\documentclass[12pt]{article}

\usepackage{graphics}
\usepackage{color}
\usepackage{amssymb}
\usepackage{latexsym}

\font\manual=manfnt \def\dbend{\lower3.5pt\hbox{\manual\char127}}

\def\ie{{\it i.e.}}

\def\cf{{\it c.f.}}

\def\etc{{\it etc}}

\def\st{\scriptstyle}
\def\sst{\scriptscriptstyle}

\def\frac#1#2{{#1\over#2}}
\def\coeff#1#2{{\textstyle{#1\over #2}}}
\def\half{\frac12}
\def\hf{{\textstyle\half}}

\def\IR{{\mathbb R}}

\def\IZ{{\mathbb Z}}

\catcode`\@=11
\def\slash#1{\mathord{\mathpalette\c@ncel{#1}}}
\def\underrel#1\over#2{\mathrel{\mathop{\kern\z@#1}\limits_{#2}}}

\catcode`\@=12
\def\sinh{{\rm sinh}} 	
\def\cosh{{\rm cosh}}

\def\exp{{\rm exp}}

\def\CC{{\cal C}} 
\def\DD{{\cal D}}

\def\GG{{\cal G}} 
\def\HH{{\cal H}}

\def\LL{{\cal L}} 
\def\MM{{\cal M}} 
\def\NN{{\cal N}} 
\def\OO{{\cal O}}

\def\TT{{\cal T}}

\def\XX{{\cal X}} 
\def\YY{{\cal Y}}

\def\unlockat{\catcode`\@=11}
\def\lockat{\catcode`\@=12}
\unlockat
\def\newsec#1{\global\advance\secno by1\message{(\the\secno. #1)}
\global\subsecno=0\global\subsubsecno=0\eqnres@t\noindent
{\bf\the\secno. #1}
\writetoca{{\secsym} {#1}}\par\nobreak\medskip\nobreak}
\global\newcount\subsecno \global\subsecno=0
\def\subsec#1{\global\advance\subsecno
by1\message{(\secsym\the\subsecno. #1)}
\ifnum\lastpenalty>9000\else\bigbreak\fi\global\subsubsecno=0
\noindent{\it\secsym\the\subsecno. #1}
\writetoca{\string\quad {\secsym\the\subsecno.} {#1}}
\par\nobreak\medskip\nobreak}
\global\newcount\subsubsecno \global\subsubsecno=0
\def\subsubsec#1{\global\advance\subsubsecno by1
\message{(\secsym\the\subsecno.\the\subsubsecno. #1)}
\ifnum\lastpenalty>9000\else\bigbreak\fi
\noindent\quad{\secsym\the\subsecno.\the\subsubsecno.}{#1}
\writetoca{\string\qquad{\secsym\the\subsecno.\the\subsubsecno.}{#1}}
\par\nobreak\medskip\nobreak}
\def\subsubseclab#1{\DefWarn#1\xdef
#1{\noexpand\hyperref{}{subsubsection}%
{\secsym\the\subsecno.\the\subsubsecno}%
{\secsym\the\subsecno.\the\subsubsecno}}%
\writedef{#1\leftbracket#1}\wrlabeL{#1=#1}}
\lockat
\newcommand{\dgr}{\dagger}

\newcommand{\txfc}[2]{{\textstyle{\frac{#1}{#2}}}}

\newcommand{\nnmb}{\nonumber}

\newcommand{\com}[2]{[\, #1\, ,\, #2\,]}
\newcommand{\acm}[2]{\{\, #1\, ,\, #2\,\}}
\newcommand{\spwd}[2]{\hspace{#1}\mbox{#2}\hspace{#1}}

\definecolor{wtmgreen}{rgb}{0,0.4,0}

\newcommand{\mc}{\mathcal}

\newcommand{\be}{\begin{equation}}
\newcommand{\ee}{\end{equation}}

\newcommand{\bbb}{\begin{eqnarray}}
\newcommand{\eee}{\end{eqnarray}}
\newcommand{\pref}[1]{(\ref{#1})}

\begin{document}

%

\def\ads{AdS}
\def\btz{{\sst BTZ}}
\def\mbar{{\bar m}}
\def\N{{\bf N}}
\def\P{{\bf P}}
\def\S{{S}}
\def\T{{T}}
\def\pp{{\bf p}}
\def\str{{\sst\rm str}}
\def\lstr{\ell_{s}}
\def\lpl{\ell_{p}}
\def\rads{\ell}


\begin{titlepage}
\rightline{EFI-01-17}

\rightline{hep-th/0106171}

\vskip 3cm
\centerline{\Large{\bf String Theory on AdS Orbifolds}}

\vskip 2cm
\centerline{
Emil J. Martinec\footnote{\texttt{e-martinec@uchicago.edu}}~~ and ~~
Will McElgin\footnote{\texttt{wmcelgin@theory.uchicago.edu}}}
\vskip 12pt
\centerline{\sl Enrico Fermi Inst. and Dept. of Physics}
\centerline{\sl University of Chicago}
\centerline{\sl 5640 S. Ellis Ave., Chicago, IL 60637, USA}

\vskip 2cm

\begin{abstract}

We consider worldsheet string theory on \(\mathbb{Z}_{N}\)
orbifolds of \(AdS_{3}\) associated with conical singularities.
If the orbifold action includes
a similar twist of \(S^{3}\), supersymmetry is preserved,
and there is a moduli space of vacua arising from
blowup modes of the orbifold singularity. We exhibit the spectrum,
including the properties of
twisted sectors and states obtained by fractional spectral flow.
A subalgebra of the spacetime superconformal symmetry
remains intact after the $\IZ_N$ quotient,
and serves as the spacetime symmetry algebra
of the orbifold.

\end{abstract}

\end{titlepage}

\newpage

\setcounter{page}{1}


\section{\label{introsec}Introduction}

The development of the AdS/CFT correspondence
(see~\cite{agmoo} for a review, and further references)
has given us a wealth of examples of dual realizations
of gravity in asymptotically anti-de Sitter
space via low energy conformal field theories.
A key property exhibited by all of these constructions
is holography: the number of degrees of freedom
describing physics in a region of space is
bounded by the area of that region in Planck units.
Unfortunately, the dual description repackages
those degrees of freedom in such a way that it is
difficult to give a concrete quasilocal description
of physics in bulk spacetime, although a number
of qualitative checks may be performed,
see for example~\cite{suswit,bdhm,bklt,peetpolch,horitz,pst}.
In particular, one would like to obtain control
of a description of black hole formation and evaporation
where one sees the horizon, can describe the experience
of the infalling observer, \etc, in order to finally
resolve the puzzles of black hole quantum mechanics.

The present investigation began as an attempt to
formulate a situation where one could approach
the formation of a black hole state beginning from
a description that was string theoretic, local in
spacetime, and (as much as possible) perturbative.
Our starting point is
the description of Giveon, Kutasov and Seiberg~\cite{GKS}
of perturbative string theory in $\ads_3\times\S^3\times\T^4$,
which is the near horizon geometry of the bound state
of $p$ fundamental strings and $k$ NS fivebranes.
The string frame metric on the globally $\ads$ spacetime is
\be\label{adsmetric}
      ds^2=k\left(-\cosh^{2}\!\rho\,dt^{2}\,+\,d\rho^{2}
	\,+\,\sinh^{2}\!\rho\,d\phi^{2}\,\right)
\ee
The relations between the string scale $\lstr$,
three-dimensional Planck scale $\lpl$, and
$\ads$ curvature scale $\rads$ of the geometry
are as follows:
\be\label{scales}
\frac\lstr{\lpl} = 4p\sqrt{k}	\quad,\qquad
\frac\rads{\lpl} = 4pk	\quad,\qquad
\frac\rads\lstr 	   = \sqrt{k}	\quad.
\ee
In addition, the six-dimensional string coupling is $g_6^2=k/p$.%
\footnote{The six-dimensional string coupling is
$g_6^2=g_s^2\lstr^4/V_4$,
where $V_4$ is the volume of $\T^4$ or K3; there is a duality
$g_6\rightarrow 1/g_6$, and a condition $V_4/\lstr^4<p/k$
for the description in terms of fundamental strings to be
valid~\cite{GKS,martsahak}.}
Introducing a pointlike object of mass below the
bound for black hole formation into this spacetime creates a conical
singularity. This may be described in terms of a global
$\ads$ geometry with an angular variable $\tilde{\phi}$ which is
identified under $\,\tilde{\phi}\sim\tilde{\phi}+2\pi\gamma\,$.
The relation between the deficit angle and the mass
in terms of the $\ads$ scale $\rads=\lstr\sqrt{k}$ is
\be\label{massdef}
      \rads M\,=\,-\hf pk\gamma^2\,=\,
      -\gamma^2\,\frac{\rads}{8\,\lpl}\ ,
\ee
with $\gamma=1$ corresponding to the global $\ads_{3}$ spacetime,
and the limit $\gamma\rightarrow0$ describing the extremal
BTZ black hole (with $M=0$ in our conventions)%
~\cite{btz,btzh}. The scale \(\lpl\) is the Planck length
which appears in the low energy gravity action defined on the quotient
spacetime. In worldsheet string theory, we know how to make
a conical singularity in the target space geometry --
take the orbifold quotient!
After laying out in section~\ref{wzwsec}
our conventions for the target space sigma models on
$\ads_3\equiv SL(2,\IR)$ and $\S^3\equiv SU(2)$, in
sections~\ref{orbsec},~\ref{suporbsec}
we describe the spectrum of the $\IZ_N$ orbifold of these
WZW models. Supersymmetry requires that one embeds
the $\IZ_N$ such that it acts simultaneously on $\ads_3$ and $S^3$.
The quantization of the $H$ flux through $S^3$ then
requires that $N$ divides the number of fivebranes $k$.%
\footnote{Similar geometries have been considered in
\cite{Coussaert:1994jp,Izquierdo:1995jz,cvelar,%
dmvw,bdkr,malmao,Mathur:2001pz,Lunin:2001fv}.}
We describe these orbifolds in section~\ref{suporbsec}.
They describe a class of conical deficit spacetimes
which one might consider as supersymmetric point objects of mass
\be\label{ptmass}
      \rads M=-\frac{pk}{2N^2}
\ee
embedded in $\ads_3$.  The $N-1$ twisted sectors
contribute moduli, the blowup modes and $B$-fluxes
of the orbifold singularity, that might be thought
of as internal excitations of the object.
When $k$ and $N$ are large, these internal excitations
might be thought of as precursors of the states of black holes.
The properties of the moduli space are briefly discussed
in section~\ref{modsec}, where we also discuss other
worldsheet CFT's related to the orbifold.

An interesting question, which we will not fully resolve,
is what the worldsheet orbifold considered here translates to as an
operation on the {\it spacetime} CFT, which for generic
values of its moduli is the sigma model
on the moduli space of instantons~\cite{stromvafa}.
The target space of the orbifold
is not invariant under $SL(2,\IR)$ isometries,
as it would be if it were a CFT vacuum.
Rather it resembles more a nontrivial state in a CFT.
We will show that the spacetime charges of the orbifold
lead one to identify it as a particular BPS state in the original
spacetime CFT.  Thus we see that, in this case, distinct worldsheet
theories (the $\ads_3\times\S^3\times\T^4$ sigma model
and its orbifold) describe perturbative string theory in
the vicinity of different states in the same nonperturbative theory.
The state corresponding to the orbifold is obtained
from the spacetime CFT vacuum by the application of
a vertex operator carrying macroscopic amounts
of spacetime energy and angular momentum.
We will also see that there are new moduli beyond the moduli
space of string theory on $\ads_3\times\S^3\times\T^4$
arising from the twisted sectors of the orbifold;
these might be interpreted as describing a
$4(N-1)$ parameter family of states having the
same global charges as the orbifold.

The orbifold quotient acts on
the $\NN=(4,4)$ superVirasoro algebra of the spacetime CFT.
Section \ref{virsec} examines the effect of the
orbifold projection on the spacetime superVirasoro symmetries.
The superconformal algebra of the theory as described on the covering
space, global $\ads_{3}\,$,
consists of a special set of analytic diffeomorphisms of
the asymptotic geometry~\cite{brohen,stromads},
whose Virasoro central charge is
$\tilde{c}=6\tilde{p}k=3\rads/2\tilde{\lpl} $;
the orbifold projection selects a certain $\IZ_N$
invariant subalgebra. Here the Planck length in the covering space is
given by $\tilde{\lpl}=N\lpl$ since the volume over which
the Einstein-Hilbert Lagrangian is integrated is $N$ times larger than
that of the quotient spacetime. Equivalently, long strings
which wrap $\tilde{p}$ times in the covering space wrap
$p=N\tilde{p}$ times in the asymptotically $\ads_{3}$
quotient space. The spacetime orbifold is a representation of
this invariant subalgebra rather than the full Virasoro
algebra. It turns out that this invariant subalgebra
is itself a twisted $\NN=(4,4)$ superVirasoro algebra,
having central charge $c=N\tilde{c}$. The twisted sector
spectrum should consist of representations of the subalgebra which
do not lift to the full superVirasoro algebra. The orbifold spacetime
is the \(-1/N\) fractional spectral flow of a Ramond vacuum state
which has \(1/N\) of the \(SU(2)\ \rm R\) charge of the
Ramond vacuum state of maximal R charge $c/12$.
The central charge of this superVirasoro
algebra is what is expected from considerations of gravity on
asymptotically \(\ads_{3}\) spacetimes. That is
$c=N\tilde{c}=3\rads/2\lpl\,$ and is the same as the
unorbifolded parent theory on global
$\ads_3\times\S^3\times\T^4$.

We also consider briefly a
second class of orbifolds which arises if we relax the
requirement of supersymmetry; then one can allow the
orbifold to act purely on $\ads_3$, and there is now no
restriction on the allowed values of $N$.
Because $N$ can be arbitrarily large, one can come arbitrarily
close to the BTZ threshold $M=0$.  Nevertheless, the spacetime
theory has tachyons in twisted sectors (of string scale mass),
indicating that the `object' is unstable.
We propose this background as a useful arena in which to explore
the dynamics of tachyon condensation in the closed string sector;
the tachyon instability is confined to the region of the
orbifold singularity rather than being spread over all of
spacetime, hence one might imagine the orbifold theory
describes an unstable point in configuration space that
then decays into a final state of strings having
the same ADM energy.  This situation is thus much like
the decay of unstable D-branes \cite{sentach},
a topic of some recent interest.
A possible obstruction to this interpretation
is the nontrivial K-theory charge carried by fractional D-branes
sitting at the orbifold singularity; this charge must
disappear (perhaps by a RR Higgs mechanism) after
tachyon condensation, or else our proposal for the final state
after tachyon condensation is incorrect.
The mass of the fractional branes will in general be
a function of the tachyon condensate, and could
vanish at some point; one could indeed then imagine
that the further decay of the system involves
condensation of branes that screens away the RR charges.
Of course, it is also possible that the system is simply unstable,
and decays violently, much as in~\cite{witkk}
(see~\cite{hormy} for a recent discussion in the context
of $\ads$ spacetimes).

Finally, we note that there may be interesting nonperturbative
generalizations of our construction.  In the moduli space
of the bound state of $p$ onebranes and $k$ fivebranes
are other perturbative limits~\cite{seibwit,larsmart} for any
other charges $p'$ and $k'$ such that $p'k'=pk$.  Each of
these perturbative limits has a GKS-type description (except $p$ or $k$
equal to one), and thus admits an orbifold by $\IZ_N$ where
$N$ divides $k'$.  It would thus appear that one can orbifold
by the cyclic group whose order is any proper divisor of $pk$.
In other perturbative limits than the ones where this
is naturally defined, the orbifold will appear as an
operation that divides by a group {\it larger} than would
naively be allowed by perturbative string theory, and yet
the resulting background must still be consistent.
If so, then the set of allowed orbifold constructions
in string theory would appear to be larger than heretofore known.


\section{\label{wzwsec}The Bosonic WZW model for \(SL(2,\mathbb{R})\)}

String theory on \(AdS_{3}\) may be formulated in terms of the
WZW model~\cite{GEPWIT} on the universal cover of the
\(SL(2,\mathbb{R})\) group manifold. The bosonic WZW action is given by
\be
S[g]\:=\:\frac{k}{8\pi}\int_{\Sigma^{2}}
d^{2}\sigma\,\delta^{\alpha\beta}\,\mbox{tr}\!
\left(\,g^{-1}\partial_{\alpha}g\,
g^{-1}\partial_{\beta}g\,\right)
\;\;+\;\;\frac{ik}{12\pi}\int_{M^{3}}\mbox{tr}\!
\left(\,\omega\wedge\omega\wedge\omega\,\right)\ .
\ee
Here \(M^{3}\) is a three manifold which has the Euclidean
worldsheet \(\Sigma^{2}\) as its boundary; \(\omega=g^{-1}dg\)
is the Maurer-Cartan form on \(SL(2,\mathbb{R})\).
This action can be expressed in the conventions
of~\cite{PSKI} as the nonlinear sigma model
\be
S[X]\:=\:\frac{k}{2\pi}\int_{\Sigma^{2}} d^{2}z \left(
G_{\mu\nu}+B_{\mu\nu}\right)\partial X^{\mu}\,\bar{\partial}X^{\nu}\ ;
\ee
where the length scale \(\rads\) of \(AdS_{3}\) that
would appear in \(G\) and \(B\)  has been absorbed into \(k\)
so that \(k=\rads^{2}/\lstr^2\,\). We parameterize \(AdS_{3}\) in
global `cylindrical' coordinates \(X^{\mu} =(t,\rho,\phi)\,\).
The generators of \(SL(2,\mathbb{R})\)
\be
\tau^{1}=\frac{i}{2}\,\sigma^{3}\hspace{0.5cm}
\tau^{2}=\frac{i}{2}\,\sigma^{1}\hspace{0.5cm}
\tau^{3}=\frac{1}{2}\,\sigma^{2}
\ee
satisfy the algebra
\be
\com{\tau^{a}}{\tau^{b}}\,=\,i\,{\epsilon^{ab}}_{c}\tau^{c}\ ,
\ee
where the metric is \(\eta^{ab}
=diag(+1,+1,-1)=-2\,\mbox{tr}(\tau^{a}\tau^{b})\)
and \(\epsilon^{123}=1\,\).
In terms of these conventions, an element of
\(SL(2,\mathbb{R})\) is given by
\be
g[X]\:=\:e^{2i\theta_{-}\tau^{3}}
e^{-2i\rho\tau^{1}}e^{2i\theta_{+}\tau^{3}}\ ,
\ee
where \(\theta_{\pm}=(t\pm\phi)/2\,\). This leads to the following
forms for \(G\) and \(B\)
\bbb
G&=&-\cosh^{2}\!\rho\,dt^{2}\,+\,d\rho^{2}
	\,+\,\sinh^{2}\!\rho\,d\phi^{2}\nnmb \\
B&=&\sinh^{2}\!\rho\,dt\wedge d\phi\ .
\label{GandB}
\eee
Taking \(\varepsilon\) to be the right handed (with respect to
\((t,\rho,\phi)\,\)) volume form on \(AdS_{3}\) this is equivalent to
\be
H\,=\,\rads^{2}\,dB\,=\,-\,\frac{2}{\rads}\,\varepsilon\ .
\ee
The WZW action is invariant under the transformation
\be
g(z,\bar{z})\,\rightarrow\,\Omega(z)g(z,\bar{z})\bar{\Omega}(\bar{z})^{-1}\ ,
\ee
leading to the currents:
\be
\mathcal{J}(z)\,=\,\mathcal{J}_{a}\tau^{a}
\,=\,-\,\frac{k}{2}\,\partial g\,g^{-1}
\quad,\qquad
\bar{\mathcal{J}}(\bar{z})\,=\,\bar{\mathcal{J}}_{a}\tau^{a}
\,=\,-\,\frac{k}{2 }\,g^{-1}\,\bar{\partial} g\ .
\ee
A modified \(\left(J^{a},\bar{J}^{a}\right)\)
definition of the currents will be used below. Introducing Cartesian
coordinates \((x^{1},x^{2},x^{3})=\sqrt{k}\,(\rho\cos\phi,\rho\sin\phi,t)\),
in the (large \(k\)) flat space limit the currents become
\be
\mathcal{J}^{a}\,\approx\,i\,\sqrt{k}\,\partial x^{a}
\hspace{2cm}
\bar{\mathcal{J}}^{a}\,\approx\,i\,\sqrt{k}\,\bar{\partial} x^{a}\ .
\label{fslcurrent}
\ee
Consider the behavior of a field \(A(z,\bar{z})\) under the
infinitesimal transformation
\be
-i\,\delta g\,=\,\epsilon(z)\,g\,-\,g\,\bar{\epsilon}(\bar{z})\ ;
\ee
the Ward identity then takes the form
\be
i\,\delta A(w,\bar{w})\,=\,\oint_w\frac{dz}{2\pi i}\,
\epsilon_{a}\mathcal{J}^{a}(z)\,A(w,\bar{w})
\:+\:\oint_w\frac{d\bar{z}}{2\pi i}\,
\bar{\epsilon}_{a}\bar{\mathcal{J}}^{a}(\bar{z})\,A(w,\bar{w})\ .
\ee
Under an infinitesimal time translation \(\epsilon=\delta t\,\tau^{3}\) and
\(\bar{\epsilon}=-\delta t\,\tau^{3}\), one finds the energy operator
\be
E_{\sst SL(2)}\:=\:\mathcal{J}^{3}_{0}\,+\,\bar{\mathcal{J}}^{3}_{0}\:=\:
\oint\frac{dz}{2\pi i}\,\mathcal{J}^{3}\:-\:
\oint\frac{d\bar{z}}{2\pi i}\,\bar{\mathcal{J}}^{3}\ .
\ee
An infinitesimal rotation \(\epsilon=-\delta\phi\,\tau^{3}\) and
\(\bar{\epsilon}=-\delta\phi\,\tau^{3}\)
yields the angular momentum
\be
L_{\sst SL(2)}\:=\:\mathcal{J}^{3}_{0}\,-\,\bar{\mathcal{J}}^{3}_{0}\ .
\ee
It turns out that the Ward identity implies that the currents
\(\left(\mathcal{J}^{a},\bar{\mathcal{J}}^{a}\right)\,\)
satisfy OPEs with structure constants of opposite sign.
This leads to the convention that \(\bar{\mathcal{J}}^{+}\) lowers
the \(\bar{\mathcal{J}}^{3}\) eigenvalue and does not conform to the
conventions for monodromies and parafermion representations in the
next two sections. This can be remedied while keeping
the same form for \(E_{\sst SL(2)}\) and \(L_{\sst SL(2)}\) in terms of
the zero modes
by defining the currents as follows
\bbb
J^{3}&=&\mathcal{J}^{3}\hspace{1cm}J^{\pm}\,=\,
J^{1}\,\pm\,i\,J^{2}\,=\,\mathcal{J}^{\pm}
\nnmb\\[0.2cm]
\bar{J}^{3}&=&\bar{\mathcal{J}}^{3}\hspace{1cm}\bar{J}^{\pm}\,=\,
\bar{J}^{1}\,\pm\,i\,\bar{J}^{2}\,=\,\bar{\mathcal{J}}^{\mp} \ .
\eee
The Ward identity then implies the OPEs
\bbb
       J^{a}(z)J^{b}(w)\,&\sim&\,
	\frac{(k/2)\eta^{ab}}{(z-w)^{2}}\,+\,
	\frac{i{\epsilon^{ab}}_{c}J^{c}(w)}{(z-w)}
\nnmb\\[0.2cm]
       \bar{J}^{a}(\bar{z})\bar{J}^{b}(\bar{w})\,&\sim&\,
	\frac{(k/2)\eta^{ab}}{(\bar{z}-\bar{w})^{2}}\,+\,
	\frac{i{\epsilon^{ab}}_{c}
	\bar{J}^{c}(\bar{w})}{(\bar{z}-\bar{w})}\ .
\label{jjope}
\eee

\subsubsection*{The unflowed \(SL(2,\mathbb{R})\) representations}
We briefly review the \(SL(2,\mathbb{R})\) representation content as
discussed for example in~\cite{DPL}.
As discussed in~\cite{M+O}, this content does not enumerate all of the
primary states of strings on \(AdS_{3}\,\). The additional
representations, which can be described by spectral flow of those
discussed in~\cite{DPL}, and their relation to the spectrum of the orbifold
will be discussed in the next section. The ``unflowed'' current
algebra primaries
of the \(SL(2,\mathbb{R})\) WZW model are arranged into three sectors:
\be
\mathcal{D}^{+}_{j}\times\mathcal{D}^{+}_{j}\hspace{0.5cm},\hspace{0.5cm}
\mathcal{D}^{-}_{j}\times\mathcal{D}^{-}_{j}\hspace{0.5cm},\hspace{0.5cm}
\mathcal{C}^{\alpha}_{j}\times\mathcal{C}^{\alpha}_{j}\hspace{0.5cm}.
\ee
Here \({\DD}^{\pm}_{j}\) are
discrete representations of the universal cover
of \(SL(2,\mathbb{R})\) which are described as follows:
\be
\DD^{\pm}_{j}\,=\,\left\{\,\left|j,m\right\rangle\,\left. \right|\:
m\,=\,\pm\left(j+n\right)\hspace{0.3cm};\hspace{0.3cm}
\left(n\geq 0\right)\in\mathbb{Z}\,\right\}
\ee
The states with \(n=0\) satisfy \(J^{\mp}_{0}\left|j,\pm j\right\rangle=0\,\).
Unitarity requires \(j\in\mathbb{R}\) with \(0\leq j<k/2\,\).
These bounds are further constrained as described
in~\cite{GKS,M+O} to \(1/2\leq j <(k-1)/2\,\).
The continuous representations \({\CC}^{\alpha}_{j}\)
are described as follows:
\be
{\CC}^{\alpha}_{j}\,=\,\left\{\,\left|j,m\right\rangle\,\left. \right|\:
m\,=\,\alpha+n\hspace{0.3cm};\hspace{0.3cm}
\left(0\leq\alpha <1\right)\in\mathbb{R}
\hspace{0.3cm};\hspace{0.3cm} n\in\mathbb{Z}\,\right\}\ .
\ee
Unitarity requires \(j=1/2+is\) where \(s\in\mathbb{R}\,\).
Note that the eigenvalue of the quadratic Casimir is bounded from above for
\(\DD^{\pm}_{j}\) by \(-j(j-1)\leq 1/4\,\). This corresponds to the
\(\rads^{2}\mu^{2}>-1\) Breitenlohner-Freedman bound
for the mass \(\mu\) appearing in the Klein-Gordon equation on
\(AdS_{3}\,\). Conversely for \({\CC}^{\alpha}_{j}\)
the quadratic Casimir is bounded from below
by \(-j(j-1)\geq 1/4\,\); which implies that the
continuous representations describe tachyonic excitations. This will not be
true of the spectral flow of these representations to be described below.


\section{\label{orbsec}The Rotational Orbifold}

\subsection{Twist Ground States}

We now consider the orbifold \(AdS_{3}/\mathbb{Z}_{N}\)
formed by a \(2\pi/N\) (\(N\in \mathbb{Z}\)) rotation.
As can be seen from the parameterization
of the group elements given above, a shift
\(\phi\rightarrow\phi +\alpha\) may be induced as follows:
\be
       g(\phi+\alpha)=e^{i\alpha\tau^{3}}g(\phi)e^{-i\alpha\tau^{3}}\ .
\ee
Thus the group field \(g(z,\bar z)\) of the WZW model has the monodromy
\be
g(e^{2\pi i}z,e^{-2\pi i}\bar{z})\,=\,
e^{2\pi iq/N\tau^{3}}g(z,\bar{z})e^{-2\pi iq/N\tau^{3}}
\ee
in the presence of the vertex operator \(\sigma_{q}\) which
creates the ground state of the \(q\in \mathbb{Z}_{N}\)
twisted sector. The holomorphic currents have monodromies
\bbb
       J^{3}(e^{2\pi i}z)&=&J^{3}(z)
\nnmb\\[0.2cm]
       J^{\pm}(e^{2\pi i}z)&=&e^{\pm 2\pi iq/N}\,J^{\pm}(z)\ .
\label{monodromy}
\eee
leading to the mode expansions for the \(q\)th twisted sector
\bbb
       J^{3}(z)&=&\sum_{n\in\mathbb{Z}}\,z^{-n-1}\,J^{3}_{n}
\nnmb\\[0.2cm]
       J^{\pm}(z)&=&\sum_{n\in\mathbb{Z}}\,
	z^{-n-1\pm q/N}\,J^{\pm}_{n\mp q/N}\ .
\label{jmod}
\eee
Similarly, the anti-holomorphic currents have the monodromies
\bbb
       \bar{J}^{3}(e^{-2\pi i}\bar{z})&=&\bar{J}^{3}(\bar{z})
\nnmb\\[0.2cm]
       \bar{J}^{\pm}(e^{-2\pi i}\bar{z})&=&
	e^{\mp 2\pi iq/N}\,\bar{J}^{\pm}(\bar{z})
\eee
and mode expansions
\bbb
       \bar{J}^{3}(\bar{z})&=&\sum_{n\in\mathbb{Z}}\,
	\bar{z}^{-n-1}\,\bar{J}^{3}_{n}
\nnmb\\[0.2cm]
       \bar{J}^{\pm}(\bar{z})&=&\sum_{n\in\mathbb{Z}}\,
	\bar{z}^{-n-1\pm q/N}\,\bar{J}^{\pm}_{n\mp q/N}\ .
\eee
Imposing the condition that \(\sigma_{q}\) is a current algebra
primary with \(E=L=0\,\) :
\be
J^{a}_{\alpha}\left|\sigma_{q}\right\rangle\,=\,0
\hspace{0.5cm}\forall\,\alpha\geq 0\spwd{1cm}{and}
\bar{J}^{a}_{\beta}\left|\sigma_{q}\right\rangle\,=\,0
\hspace{0.5cm}\forall\,\beta\geq 0\ ,
\ee
the mode algebras implied by the OPEs of the currents may
be used to compute the dimension of the twist vertex operators
\be
h(\sigma_{q})=\frac{1}{2\pi i}\oint dz\, z \left\langle\sigma_{q}\right|
T(z)\left|\sigma_{q}\right\rangle\ .
\ee
With \(T(z)\) given by the Sugawara construction
$T(z)=\frac{1}{k-2} \eta_{ab}J^{a}\! J^{b}(z)$,
one finds
\be
h(\sigma_{q})\,=\,\bar{h}(\sigma_{q})\,=\,\frac{k}{k-2}\,
\frac{1}{2}\,\left(q/N\right)\left(1-q/N\right)\ .
\label{twdim}
\ee
Note that in the flat space limit
$k\rightarrow\infty$, the currents
become the translation currents~\pref{fslcurrent}
with the $\IZ_N$ monodromies~\pref{monodromy}.
The properties of the twist operators agree with those
computed in~\cite{DFMS}\,.


\subsection{\(SL(2,\mathbb{R})\) Parafermions}

A more explicit construction for the twist
vertex operators \(\sigma_{q}\) can be given
in terms of parafermions.
We first introduce a parafermion representation for the currents
\be
J^{3}\,=\,-\,\sqrt{k/2}\,\partial{X}\hspace{1cm},\hspace{1cm}
J^{\pm}\,=\,\psi^{\pm}\,e^{\pm\sqrt{2/k}\,{X}}\ .
\label{pfrepsl2}
\ee
And similarly for the anti-holomorphic currents.
Here \({X}\) is a holomorphic field with OPE
\be
{X}(z)\,{X}(w)\,\sim\,-\ln{(z-w)}\ .
\label{xxope}
\ee
The parafermions \(\psi^{\pm}\) have OPEs
\be
\psi ^{+}(z)\, \psi^{-}(w)\: \sim \:
k\;(z-w)^{-2-2/k}\quad,\qquad
\psi ^{\pm }(z)\, \psi ^{\pm }(w)\: \sim \: 0\quad,
\ee
which reflect the OPEs for the \(SL(2,\mathbb{R})\) currents~\pref{jjope}.
The current algebra primaries for the \(SL(2,\mathbb{R})\) WZW model
are given in terms of parafermions by
\be
\Phi^{\sst SL(2)}_{jm\bar{m}}\,=\,\Psi^{\sst SL(2)}_{jm\bar{m}}\,
e^{\sqrt{2/k}\,\left(m{X}+\bar{m}\bar{{X}}\right)}\ .
\label{pfdef}
\ee
Here \(\left(m,\bar{m}\right)\) are the eigenvalues of the zero modes
\(\left(J^{3}_{0},\bar{J}^{3}_{0}\right)\,\) and \(j\) is the
spin eigenvalue of \(SL(2,\mathbb{R})\,\). Note that, since we are
considering the non-compact covering group of \(SL(2,\mathbb{R})\), \(j\)
is not quantized. Also note that, due to modular invariance and the
fact that the covering space is simply connected, the holomorphic
and anti-holomorphic Casimirs are equal
\be
\eta_{ab}J^{a}_{0}J^{b}_{0}\,=\,
\eta_{ab}\bar{J}^{a}_{0}\bar{J}^{b}_{0}\,=\,-j\left(j-1\right)\ .
\ee
This implies, given details of the representation content,
that \(m-\bar{m}\in\mathbb{Z}\,\). Of course this is just a consequence
of requiring that the wave function be single valued under a rotation
by \(2\pi\,\).
The dimensions of the primary fields are
\be
h\left(\Phi^{\sst SL(2)}_{jm\bar{m}}\right)\,=\,
\bar{h}\left(\Phi^{\sst SL(2)}_{jm\bar{m}}\right)\,=\,
\frac{-j\left(j-1\right)}{k-2}\ ,
\ee
hence the dimensions for the parafermions are
\be
h\left(\Psi^{\sst SL(2)}_{jm\bar{m}}\right)\,=\,
\frac{-j\left(j-1\right)}{k-2}\,+\,\frac{m^{2}}{k}\ ,
\label{pfdim}
\ee
and similarly for \(\bar{h}\) with \(m\rightarrow\bar{m}\,\).

\subsubsection*{The parafermion twist operators}
We now describe the $\IZ_N$
twist operators in terms of the parafermions. It
may be verified that, in terms of conformal dimensions and monodromies
with the currents, all of the properties of the twist vertex operator
are satisfied by
\be
\sigma_{q}\,=\,\Psi^{\sst SL(2)}_{j_{q},j_{q},j_{q}}
\spwd{1cm}{where}j_{q}\,=\,\frac{kq}{2N}\ .
\ee
Thus \(\sigma_{q}\) is the parafermion for the
lowest weight state in \(\DD^{+}_{j_{q}}\,\). In particular
using~\pref{pfdim}, as found in~\pref{twdim} above,
\be
h\left(\Psi^{\sst SL(2)}_{j_{q},j_{q},j_{q}}\right)\,=\,
\frac{kj_q-2j_q^2}{k(k-2)}\,=\,
\frac{k}{k-2}\,\frac{1}{2}\,\left(q/N\right)\left(1-q/N\right)\ .
\label{slpfdim}
\ee
The monodromies with respect to the currents may be verified by
looking at the OPEs with the primaries
\be
J^{\pm}(z)\,\Phi^{\sst SL(2)}_{j,m,\bar{m}}(w)\,\sim\,
(m\pm j)\left(z-w\right)^{-1}\Phi^{\sst SL(2)}_{j,m\pm 1,\bar{m}}(w)
\ee
which implies, through the free OPE~\pref{xxope},
\bbb
J^{+}(z)\,\Psi^{\sst SL(2)}_{j_{q},j_{q},j_{q}}(w)&\sim&
\left(z-w\right)^{-(1-q/N)}\widehat{O}^{+}(w)
\nnmb \\[0.2cm]
J^{-}(z)\,\Psi^{\sst SL(2)}_{j_{q},j_{q},j_{q}}(w)&\sim&
\left(z-w\right)^{-q/N}\widehat{O}^{-}(w)
\eee
for some vertex operators \(\widehat{O}^{\pm}\,\). This may be compared
with the result that follows from the mode expansions~\pref{jmod}\,:
\bbb
J^{+}(z)\left|\sigma_{q}\right\rangle&\sim&
z^{-(1-q/N)}\,J^{+}_{-q/N}\left|\sigma_{q}\right\rangle
\nnmb \\[0.2cm]
J^{-}(z)\left|\sigma_{q}\right\rangle&\sim&
z^{-q/N}\,J^{-}_{-(1-q/N)}\left|\sigma_{q}\right\rangle\ .
\eee
Finally, the \(J^{3}\) current is regular with respect to both
\(\sigma_{q}\) and the parafermions.


\subsection{Fractional Spectral Flow}

\subsubsection*{The integer spectral flow operator}
As described above, for the discrete representations
the \(SL(2,\mathbb{R})\) spin quantum number \(j\)
is constrained to the values \(1/2\leq j <(k-1)/2\,\). This implies,
through the mass shell condition, a limit on the dimensions of vertex operators
associated with the spaces with which \(AdS_{3}\) may be combined to create
a critical string theory. Furthermore there is the expectation that strings of
arbitrary level number should be permitted in \(AdS_{3}\,\). For
instance, a classical description of string propagation in
\(AdS_{3}\,\) involves strings that propagate to spatial infinity. These
should be associated with a continuous spectrum of physical states. These
problems and the problem that the descendants of the above primaries do
not lead to a modular invariant partition function were resolved relatively
recently in~\cite{M+O}, by considering the role of spectral
flow in $SL(2,\IR)$.  Spectral flow by $w$ units introduces
a shift in the worldsheet quantum numbers
\bbb
     L_0 &\rightarrow& L_0-J_0^3\,w-\frac k4\,w^2 \nnmb\\
     J_0^3 &\rightarrow& J_0^3+\frac k2\,w\ ,
\label{spfl}
\eee
and similarly for $\bar L_0$, $\bar J_0^3$.
In the parafermion representation, the operator that
implements this spectral flow is given by%
\footnote{In \cite{AGS}, the spectral flow operators are
referred to as twist operators; we prefer to refer
to them as spectral flow operators, reserving the term
twist operators to refer to the twisted sector vertex operators
of the $\IZ_N$ orbifold.}
\be
t_{w}\,=\,e^{w\,\sqrt{k/2}\,\left({X}+\bar{{X}}\right)}
\spwd{1cm}{where}w\in\mathbb{Z}\quad .
\label{bspecflow}
\ee
The result of the introduction of these operators is to extend the
primary states to include all vertex operators of the form
\be
\Phi^{\sst SL(2)}_{wjm\bar{m}}\,=\,\Psi^{\sst SL(2)}_{jm\bar{m}}\,
e^{\sqrt{2/k}\,\left(\left(m+\frac{k}{2}w\right){X}+
\left(\bar{m}+\frac{k}{2}w\right)\bar{{X}}\right)}\quad .
\ee
This leads to the emergence of discrete states of arbitrary level
number as well as continuous states that are not tachyonic.

\subsubsection*{The fractional spectral flow operator}
The spectrum of the orbifold consists of vertex operators which
comprise a closed mutually local OPE algebra which includes
the twist vertex operators.
For the primary states this translates to
\be
\exp\left(i\left(2\pi /N\right)L_{\sst SL(2)}\right)
\left|\Phi^{\sst SL(2)}_{wjm\bar{m}}\right\rangle\,=\,
\left|\Phi^{\sst SL(2)}_{wjm\bar{m}}\right\rangle
\ee
which implies the condition \(m-\bar{m}\in N\mathbb{Z}\,\). The twisted
sectors consist of states resulting from the OPEs of the
surviving untwisted sector states with the twist vertex operators.
Consider the most singular term in the OPE of an unflowed lowest
weight primary of spin \(j=j_{q}\) with the twist operator of the
\(\left(N-q\right)\) sector
\bbb
\lefteqn{\Phi^{\sst SL(2)}_{0j_{q}j_{q}j_{q}}\left(z,\bar{z}\right)\,
	\sigma_{N-q}\left(w,\bar{w}\right)}\hspace{1cm} \nnmb
	\\[0.2cm]  & = &
		e^{q/\!{\sst N}\,\sqrt{k/2}\,\left({X}(z)+\bar{{X}}
		(\bar{z})\right)}\, \sigma_{q}\left(z,\bar{z}\right)\,
		\sigma_{N-q}\left(w,\bar{w}\right) \nnmb
	\\[0.2cm] & \sim &
		\frac{C_{q,N-q}}{|z-w|^{4h(\sigma_{q})}}\;
		\sigma_{N}\left(w,\bar{w}\right)\,
		e^{q/\!{\sst N}\,\sqrt{k/2}\,
		\left({X}(w)+\bar{{X}}(\bar{w})\right)}\,+\,\ldots\ .\
\eee
Note that \(\sigma_{N}\) has dimension zero
\be
h\left(\sigma_{N}\right)\,=\,
h\left(\Psi^{\sst SL(2)}_{\frac{k}{2},
\frac{k}{2},\frac{k}{2}}\right)\,=\,0\ .
\ee
It is natural to associate it with the identity operator. We have
thus introduced a fractional spectral flow operator into the spectrum
\be
t_{q/\!{\sst N}}\,=\,e^{q/\!{\sst N}\,\sqrt{k/2}\,\left({X}+\bar{{X}}\right)}
\spwd{1cm}{where}q\in\mathbb{Z}_{N} \ .
\ee

\begin{figure}[h]
\label{figone}
\begin{center}
\[
\mbox{\begin{picture}(102,200)(0,0)
\includegraphics{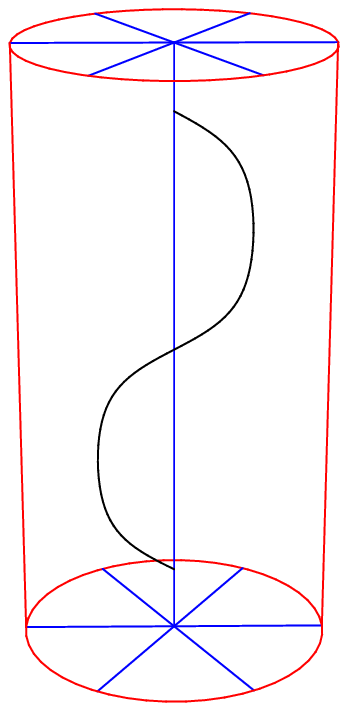}
\put(20,96){\(t_{q/\!{\sst N}}\)}
\put(20,80){\(\Longrightarrow\)}
\end{picture}\hspace{2cm}
\begin{picture}(102,200)(0,0)
\includegraphics{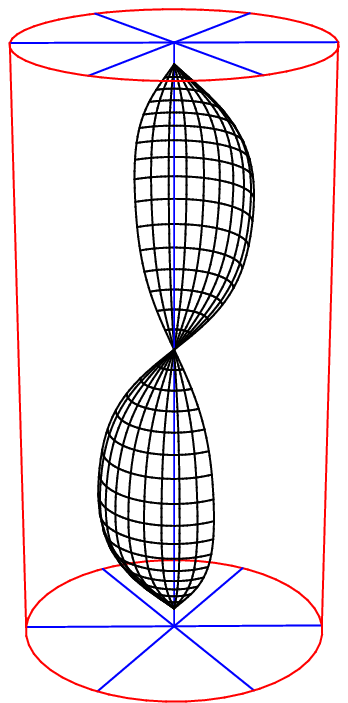}
\end{picture}}
\]
\caption{Fractional spectral flow of a timelike geodesic producing a
string in the \(q=2\) twisted sector for the \(N=6\) orbifold.}
\end{center}
\end{figure}

Fractional spectral flow
generates primary states of the orbifold theory that are of the form
\be
\Phi^{\sst SL(2)/\mathbb{Z}_{N}}_{pjm\bar{m}}\,=\,\Psi^{\sst
SL(2)}_{jm\bar{m}}\,
e^{\sqrt{2/k}\,\left(\left(m+\frac{k}{2}\frac{p}{N}\right){X}+
\left(\bar{m}+\frac{k}{2}\frac{p}{N}\right)\bar{{X}}\right)}\ .
\ee
Here \(m-\bar{m}\in N\mathbb{Z}\) and \(p\in \mathbb{Z}\,\). The
twisted sector of the primary is given by \(q=N-(p\,\mbox{mod}\,N)\) as
may be verified by the monodromy with the currents.
The unflowed primary states, being representations of the zero mode
algebra, correspond in the classical limit to geodesics of the
geometry. As explained in~\cite{M+O} the geodesics corresponding to
the discrete representations of the WZW model are timelike
and those of the  continuous representations are spacelike. Spectral
flow of a geodesic which passes through the origin by
\(w\in\mathbb{Z}\) stretches the geodesic in the timelike
direction, allowing spacelike geodesics corresponding to tachyonic
primaries to become timelike, and forms a string worldsheet wrapped
\(w\) times around the origin as a surface of revolution of the stretched
geodesic. The operation of fractional spectral flow
introduces strings of this type that wind a fractional number of
times, and are closed only by virtue of the orbifold identification.
The example of fractional spectral flow on a
timelike geodesic for \(N=6\,\), \(w=0\) and \(q=2\) is shown in
figure~\ref{figone}.

\section{\label{suporbsec}The
\(\left(AdS_{3}\times S^{3}\right)\!/\mathbb{Z}_{N}\) Orbifold}

\subsection{Bosonic Structure}

To construct a critical bosonic string theory (\(c=26\)) that has a well
defined flat space limit, the WZW model describing string theory
on \(AdS_{3}\) must be combined with another CFT.
Here we will consider some aspects of
the bosonic string theory on a \(\mathbb{Z}_{N}\) orbifold of
\(AdS_{3}\times S^{3}\times \mathcal{N}\) associated with a combined
rotation of \(AdS_{3}\) and \(S^{3}\). Here \(\mathcal{N}\) is
the target space of a bosonic CFT which has central charge
\(c_{\sst\mathcal{N}} =20\,\). String theory on \(S^{3}\) may be
described in terms of a WZW model on the \(SU(2)\) group manifold.
The description of this WZW model is very similar to that provided
above for \(SL(2,\mathbb{R})\,\). One distinction is in the central
charges of the bosonic theories
\be
c_{\sst SL(2)}\,=\,\frac{3\,k_{\sst SL(2)}}{k_{\sst SL(2)}-2}
\hspace{1cm},\hspace{1cm}c_{\sst SU(2)}\,=\,
\frac{3\,k_{\sst SU(2)}}{k_{\sst SU(2)}+2}\ .
\ee
This arises from the difference in the value of the quadratic
Casimirs of the adjoint representations of the respective groups.
Here we will consider only \(c_{\sst SL(2)}+c_{\sst SU(2)}=6\) for
the bosonic theory, so that $k_{\sst SL(2)}-2=k_{\sst SU(2)}+2$.
The \(SU(2)\) group manifold may be parameterized by
three Euler angles. One of these angles, which will be
denoted \(\phi_{\sst SU(2)}\,\), may be shifted by the transformation
\be
       g(\phi_{\sst SU(2)}+\alpha)\,=\,e^{i\alpha\sigma^{3}/2}\,
       g(\phi_{\sst SU(2)})\,e^{-i\alpha\sigma^{3}/2}\ .
\label{phidef}
\ee
The orbifold we would like to consider identifies
\(AdS_{3}\times S^{3}\) as follows
\be
\left(\,\phi_{\sst SL(2)}\, ,\,\phi_{\sst SU(2)}\,\right)\:\sim\:
\left(\,\phi_{\sst SL(2)}+2\pi/N\, ,\,
\phi_{\sst SU(2)}-2\pi/N\,\right)\ .
\label{orbident}
\ee
The twisted sector states on this target space may be described
by considering the twisted spectra on the rotational orbifolds of the two
group manifolds separately and then pairing the \(q\) twisted sector of
\(SL(2,\mathbb{R})/\mathbb{Z}_{N}\) with the \(N-q\) twisted sector of
\(SU(2)/\mathbb{Z}_{N}\,\).

\subsubsection*{The \(SU(2)/\mathbb{Z}_{N}\) CFT}
We briefly review the \(SU(2)/\mathbb{Z}_{N}\) theory~\cite{gepqiu}.
An asymmetric version of this orbifold is described in~\cite{KLL}. When the
level \(k\) and spin \(j\) appear in this paragraph they correspond to
\(k_{\sst SU(2)}\) and \(j_{\sst SU(2)}\,\).  The current algebra
primaries of the \(SU(2)\) WZW model are expressed in terms of
parafermions as
\be
\Phi^{\sst SU(2)}_{jm\bar{m}}\,=\,\Psi^{\sst SU(2)}_{jm\bar{m}}\,
e^{i\sqrt{2/k}\,\left(m{Y}+\bar{m}\bar{{Y}}\right)}\ .
\ee
Here \(Y\) is a free boson associated with the holomorphic current
\(J_{\sst SU(2)}^{3}\,\):
\be
J_{\sst SU(2)}^{3}\,=\,i\,\sqrt{k/2}\,\partial{Y}\ .
\label{pfrepsu2}
\ee
And similary for \(\bar{Y}\,\). As for the \(SL(2,\mathbb{R})\)
case, the holomorphic and antiholomorhic Casimirs are equal. The dimension
of the primaries, as computed using the Sugawara stress tensor
$T(z)=\frac{1}{k+2}\; \delta_{ab} J^{a}\!J^{b}(z)$,
is given by
$h\left(\Phi^{\sst SU(2)}_{jm\bar{m}}\right)=
\frac{j\left(j+1\right)}{k+2}$;
subtracting the contribution of the free boson yields
the dimensions of the parafermions
\be
h\left(\Psi^{\sst SU(2)}_{jm\bar{m}}\right)\,=\,
\frac{j\left(j+1\right)}{k+2}\,-\,\frac{m^{2}}{k}\ .
\ee
And similarly for \(\bar{h}\) with \(m\rightarrow\bar{m}\,\). The
twist vertex operators of the \(SU(2)/\mathbb{Z}_{N}\,\) orbifold
are given by the lowest weight state in the
\(\mathcal{D}^{\sst SU(2)}_{j_{q}}\) representation. That is
\be
\sigma_{q}^{\sst SU(2)}\,=\,\Psi^{\sst SU(2)}_{j_{q},-j_{q},-j_{q}}
\spwd{1cm}{where}j_{q}\,=\,\frac{k\,q}{2\,N}\ .
\ee
Note that the \(j=j_{q}\) representation exists since the level
\(k\) of the orbifold theory is restricted to
\(k_{\sst SU(2)}\in N\mathbb{Z}\);
\(SU(2)/\mathbb{Z}_{N}\) is a compact manifold with a volume
that is \(1/N\) of that of \(SU(2)\), and the $H$
flux threading it must be an integer.  Also note that, as for the
\(SL(2,\mathbb{R})/\mathbb{Z}_{N}\) orbifold,
\(m-\bar{m}\in N\mathbb{Z}\,\). The dimensions of the
twist operators are thus given by
\be
h\left(\sigma_{q}^{\sst SU(2)}\right)\,=\,
\frac{kj_q-2j_q^2}{k(k+2)}\,=\,
\frac{k}{k+2}\,\frac{1}{2}\,\left(q/N\right)\left(1-q/N\right)
\label{supfdim}
\ee
which, as for \(SL(2,\mathbb{R})/\mathbb{Z}_{N}\)
above, agrees with the calculation~\cite{DFMS} in the
\(k\rightarrow\infty\) flat space limit. The twist operators
imply the existence of fractional spectral flow operators
\be
t^{\sst SU(2)}_{q/\!{\sst N}}\,=\,e^{i\,q/\!{\sst N}\,
\sqrt{k/2}\,\left({Y}+\bar{{Y}}\right)}
\spwd{1cm}{where}q\in\mathbb{Z}_{N} \ .
\ee
However there is no independent spectral flow quantum number
\(w\) characterizing the representations of
the (compact) \(SU(2)\) WZW model, since integer spectral
flow maps the current algebra representations into themselves.


\subsection{The Supersymmetric Orbifold}

\subsubsection*{The supersymmetric WZW model}

We now consider the superstring on the orbifold
\(\left(AdS_{3}\times S^{3}\right)\!/\mathbb{Z}_{N}\,
\times\mathcal{N}\,\). To form a critical \(c=15\) theory, the
CFT on \(\mathcal{N}\) is required to have \(c_{\mathcal{N}}=6\,\).
Only the case \(\mathcal{N}=T^{4}\) will be described in detail;
the generalization is straightforward.
The OPEs and Sugawara stress tensor are described in
what follows for both the \(SU(2)\) and \(SL(2,\mathbb{R})\)
supersymmetric level \(k\) WZW model~\cite{DKPR} with the Killing metric
\(g_{ab}\) and quadratic Casimir \(Q\,\). The WZW supercurrent
\(C^{a}\) is expressed in terms of the total current \(J^{a}\)
and the fermions \(\psi^{a}\) as
\be
C^{a}\,=\,\psi^{a}\,+\,\theta\, J^{a}\ \ .
\ee
Here \(\theta\) is the holomorphic worldsheet Grassmann coordinate.

\(J^{a}\) and \(\psi^{a}\) satisfy the OPEs :
\bbb
J^{a}(z)J^{b}(w)&\sim&\frac{(k/2)\,g^{ab}}{(z-w)^{2}}+
\frac{i{\epsilon^{ab}}_{c}\,J^{c}(w)}{(z-w)} \nnmb \\[0.2cm]
J^{a}(z)\psi^{b}(w)&\sim&
\frac{i{\epsilon^{ab}}_{c}\,\psi^{c}(w)}{(z-w)} \\[0.2cm]
\psi^{a}(z)\psi^{b}(w)&\sim&\frac{(k/2)\,g^{ab}}{(z-w)} \nnmb
\eee
Subtracting the contribution to \(J^{a}\) which comes from the
fermionic piece of the SUSY WZW action produces the bosonic current
\be
j^{a}=J^{a}+(i/k)\,{\epsilon^{a}}_{bc}\,\psi^{b}\psi^{c}\ ,
\ee
which leads to the OPEs
\bbb
j^{a}(z)j^{b}(w)&\sim&\frac{(\tilde{k}/2)\,g^{ab}}{(z-w)^{2}}+
\frac{i{\epsilon^{ab}}_{c}\,j^{c}(w)}{(z-w)} \nnmb \\[0.2cm]
j^{a}(z)\psi^{b}(w)&\sim& 0 \ .
\eee
Where \(\tilde{k}=k-Q\,\).
The stress tensor may be obtained as usual:
\be
T\,=\,\frac{1}{\tilde{k}+Q}\,j^{a}j^{b}\,g_{ab}\,-\,
\frac{1}{k}\,\psi^{a}\partial\psi^{b}\,g_{ab}\ ,
\ee
and the Virasoro central charge is
\be
c\,=\,\frac{3\,\tilde{k}}{\tilde{k}+Q}\,+\,\frac{3}{2}\ .
\ee
The Virasoro supercurrent is given by
\be
G\,=\,\frac{2}{k}\,\left(\,g_{ab}\psi^{a}j^{b}\,-\,
\frac{i}{3k}\,\epsilon_{abc}\psi^{a}\psi^{b}\psi^{c}\,\right)\ .
\label{supg}
\ee
%
%
\subsubsection*{The superstring on \(AdS_{3}\times S^{3}\times T^{4}\)}
We briefly review here the description of the superstring on
\(AdS_{3}\times S^{3}\times T^{4}\) as it appears
in~\cite{GKS,KLL}. The fermions and the total and bosonic currents
associated with the \(\widehat{SL}(2,\mathbb{R})\) current algebra
will be denoted by \(\left(\psi^{A},J^{A},j^{A}\right)\) respectively.
For \(\widehat{SU}(2)\) they will be denoted by
\(\left(\chi^{a},K^{a},k^{a}\right)\,\). The (canonically normalized)
\(\widehat{U}(1)^{4}\) fermions and current will be denoted
\(\lambda^{j}\) and \(i\partial F^{j}\,\).
Note that the levels of the associated bosonic WZW current algebras are
shifted as described above. That is, for the superstring
\be
c_{\sst SL(2)}\,=\,\frac{3\,(k_{\sst SL(2)}+2)}{k_{\sst SL(2)}}
+\frac{3}{2}\spwd{1cm}{,}c_{\sst SU(2)}\,=\,
\frac{3\,(k_{\sst SU(2)}-2)}{k_{\sst SU(2)}}+\frac{3}{2}\ .
\ee
The condition \(c=15\) then leads to
\(k=k_{\sst SL(2)}=k_{\sst SU(2)}\,\).
The ten fermions may be
written in terms of five canonically normalized free bosons
\(H_{I}\) where \(I\in(1,\ldots ,5)\) as follows
\bbb
      i\partial H_{1} & = & -i\,\txfc{2}{k}\,
	\psi^{1}\psi^{2}\,=\,J^{3}-j^{3}   \nnmb \\[0.1cm]
      i\partial H_{2} & = & -i\,\txfc{2}{k}\,
	\chi^{1}\chi^{2}\,=\,K^{3}-k^{3} \nnmb \\[0.1cm]
      i\partial H_{3} & = & -\,\txfc{2}{k}\,\psi^{3}\chi^{3}
\label{oten}\\[0.1cm]
      i\partial H_{4} & = & -i\,\lambda^{1}\lambda^{2} \nnmb \\[0.1cm]
      i\partial H_{5} & = & -i\,\lambda^{3}\lambda^{4} \nnmb\ .
\eee
The spacetime supercharges are constructed as in~\cite{FMS} :
\be
\label{superqs}
Q_{\alpha}\,=\,\oint \frac{dz}{2\pi i}\,e^{-\varphi/2}\,S_{\alpha}(z)\ ,
\ee
where \(\varphi\) is a boson of the \(\beta ,\gamma\) superghost system and
\be
S_{\alpha}\,=\,\exp\left(\,\txfc{i}{2}\,\epsilon_{I}H_{I}\,\right)
\ee
is one of the 32 spin fields possible before the imposition of the GSO
projection and the condition of BRST invariance. Here
\(\alpha=(\epsilon_{1}\ldots\epsilon_{5})\) where \(\epsilon_{I}=\pm 1\,\).
The GSO projection amounts to imposing
\be
\prod_{I=1}^{5}\,\epsilon_{I}\,=\,1\ .
\ee
BRST invariance is not guaranteed due to the presence of the term
cubic in the fermions in $G(z)$~\pref{supg}. The cancellation of the term
of order \(z^{-3/2}\) in the \(G(z)S_{\alpha}(0)\) OPE leads to the
condition
\be
\prod_{I=1}^{3}\,\epsilon_{I}\,=\,1\ .
\label{brst}
\ee
Thus there are 8 `left-moving'
supercharges on \(AdS_{3}\times S^{3}\times T^{4}\)
as compared to 16 in flat space (and 8 more from the
right-movers).

\subsubsection*{Massless and spacetime chiral states}
As shown explicity in~\cite{GKS}, the worldsheet
\(\widehat{SL}(2,\mathbb{R})\times\widehat{SU}(2)\times\widehat{U}(1)^{4}\)
current algebra is associated with an \(\mc{N}=4\) superconformal algebra
in spacetime. The global modes of the spacetime Virasoro and current
algebras are just the charges of the associated total worldsheet currents.
Thus for \(\widehat{SL}(2,\mathbb{R})\) and
\(\widehat{SU}(2)\) :
\bbb
\LL_{0}\,=\,\oint\frac{dz}{2\pi i}\,J^{3}(z)\hspace{2cm}
\LL_{\pm}\,=\,\oint\frac{dz}{2\pi i}\,J^{\mp}(z)
\eee
and
\bbb
\TT^{a}_{0}\,=\,\oint\frac{dz}{2\pi i}\,K^{a}(z)\ .
\eee
Chiral primary states of this spacetime algebra
were described in~\cite{GKS}~\cite{KLL} in terms of the unflowed
worldsheet current algebra primaries as follows. The worldsheet
primaries of the WZW supercurrents are given by the primaries of
the bosonic (\,level \((k+2)\) for \(SL(2,\mathbb{R})\) and \((k-2)\) for
\(SU(2)\,\)) WZW model
\be
\Phi^{\sst SL(2)}_{jm\bar{m}}\:\Phi^{\sst
SU(2)}_{j'm'\bar{m}'}\:e^{i\pp\cdot F+i\bar{\pp}\cdot\bar{F}}\ \ .
\ee
Here \((\pp,\bar{\pp})\) is a vector in an even, self-dual Narain
lattice \(\Gamma^{4,4}\,\). Physical states consist of bosonic
current algebra descendants and fermionic excitations of these
primaries which are in the BRST cohomology. Suppressing the
anti-holomorphic quantum numbers, the \(\widehat{SL}(2,\mathbb{R})\)
primary states satisfy~\cite{GKS}
\be
\com{\LL_{n}}{\Phi^{\sst SL(2)}_{j,m}}\,=\,
\left(n(j-1)-m\right)\,\Phi^{\sst SL(2)}_{j,m+n}
\ee
where \(\LL_{n}\) are generators of the spacetime Virasoro algebra.
Thus these primaries can be seen to be modes of an operator
with spacetime scaling dimension \(h=j\,\).
Unitarity of the spacetime superconformal algebra requires that
\(h\geq j_{\sst SU(2)}\,\), where \(j_{\sst SU(2)}\) is the spin
associated with the spacetime \(\widehat{SU}(2)\) current algebra
\(\TT^{a}_{n}\,\). This is, of course,
also the spin of the total worldsheet \(\widehat{SU}(2)\)
current \(K^{a}\,\). The chiral primary operators in the NS sector
saturate this bound and satisfy
\(\pp=0\) and \(\N=1/2\,\), where \(\N\) is the
total level of the oscillator excitations. Physical states
which are spacetime chiral primaries correspond to massless excitations
as can be seen from the NS sector mass-shell condition
\be
\frac{-j(j-1)}{k}\,+\,\frac{j'(j'+1)}{k}\,+\,\frac{\pp\cdot
\pp}{2}\,+\,\N\,=\,\frac{1}{2}\ .
\ee
This implies \(j'=j-1\,\).%
\footnote{Note that a different convention is used
here for the \(SL(2,\mathbb{R})\) \(j\) quantum number than is used
in~\cite{GKS,KLL}.}
There are eight massless physical NS states
for a given \(j\) as well as \(m,m'\) (which are suppressed)
\bbb
      \mathcal{V}^{i}_{j} & = &
	e^{-\varphi}\,\lambda^{i}\,\Phi^{\sst SL(2)}_{j}\,
	\Phi^{\sst SU(2)}_{j'}
\nonumber\\[0.2cm]
      \mathcal{W}^{\pm}_{j} & = & e^{-\varphi}\,
	\left[\psi\,\Phi^{\sst SL(2)}_{j}\right]_{j\pm 1}\,
	\Phi^{\sst SU(2)}_{j'}
\label{untwistops}\\[0.2cm]
      \mathcal{X}^{\pm}_{j} & = & e^{-\varphi}\,
	\Phi^{\sst SL(2)}_{j}\,
	\left[\chi\,\Phi^{\sst SU(2)}_{j'}\right]_{j'\pm 1}
\nonumber
\eee
Here the quantum numbers outside the brackets refer to
\(j_{\sst SL(2)}\) and \(j_{\sst SU(2)}\,\), the spins
associated with the quadratic Casimirs of the total currents
\(J^{a}\) and \(K^{a}\) respectively. Details of the construction
of these states may be found in~\cite{KLL}. The additional two states
\bbb
     \mathcal{W}^{0}_{j} & = & e^{-\varphi}\,
	\left[\psi\,\Phi^{\sst SL(2)}_{j}\right]_{j}\,
	\Phi^{\sst SU(2)}_{j'}
\nonumber\\[0.2cm]
     \mathcal{X}^{0}_{j} & = & e^{-\varphi}\,
	\Phi^{\sst SL(2)}_{j}\,
	\left[\chi\,\Phi^{\sst SU(2)}_{j'}\right]_{j'}
\eee
are not in the BRST cohomology. Of these massless states only
\(\mathcal{W}^{-}_{j}\) and \(\mathcal{X}^{+}_{j}\) are spacetime chiral
primaries, that is are modes of spacetime operators which satisfy
\(h=j_{\sst SL(2)}=j_{\sst SU(2)}\,\). Ramond sector states are
found by applying the spacetime supercharges to these NS
states~\cite{FMS}.

In all of these vertex operators, the current algebra primaries
$\Phi^{\sst SU(2)}_{j'm'\mbar'}$ and $\Phi^{\sst SL(2)}_{jm\mbar}$
span the space of wavefunctions on $\ads_3\times S^3$
for spins less than $k/2$,
while the fermions $\psi$, $\chi$, $\lambda$ carry the
polarizations of the various supergravity modes.

%
%
\subsubsection*{Superparafermions}
Due to the contribution of the fermions to the total currents $J^a$
and $K^a$ of $SL(2,\IR)$ and $SU(2)$, respectively,
it is convenient to write the current algebra primaries in terms of
superparafermions times exponentials of the
bosons related to the total currents $J^3$ and $K^3$.
For example for the \(\widehat{SU}(2)\) superparafermions
\be
      {\Phi^{\sst SU(2)}_{jm\mbar}} \,=\,
	{\widehat{\Psi}}^{\sst SU(2)}_{jm\mbar}\;
            \exp\Bigl[i\sqrt{\txfc 2k}
	\Bigl(m\YY+\mbar\bar \YY\Bigr)\Bigr]\ \ .
\ee
Where we have bosonized the total current as
\(K^3=i\sqrt{{\st k/2}}\,\partial\YY\,\).
The bosonic current $k^3=i\sqrt{{\st\tilde k/2}}\,\partial Y$
of the parafermion construction of~\pref{pfrepsu2} (referred to
as \(J^{3}_{SU(2)}\) there)
and the boson $H_2$ of~\pref{oten} are then rewritten
in terms of $\YY$ and another boson $\HH_2$
(both canonically normalized) via%
\footnote{Recall our notation \(\tilde{k}=k-Q\,\).  Thus,
for $SU(2)$, $Q=2$ and $\tilde k=k-2$; while for $SL(2,\IR)$,
$Q=-2$ and $\tilde k=k+2$.}
\bbb
       Y &=& \sqrt{\txfc{\tilde{k}}{k}}\:\YY\,-\,
       \sqrt{\txfc{2}{k}}\:\HH_2\ \nnmb\\[0.2cm]
       H_2 &=& \sqrt{\txfc{\tilde{k}}{k}}\:
       \HH_2\,+\,\sqrt{\txfc{2}{k}}\:\YY\ \ .
\eee
One also has the following relation between the level \(k\) superparafermions
and the parafermions of the bosonic \(SU(2)\) level \(k-2\) WZW model
\be
       {\widehat{\Psi}}^{\sst SU(2)}_{jm\mbar} \,=\,
	\Psi^{\sst SU(2)}_{jm\mbar}\;
            \exp\Bigl[-i\sqrt{\txfc{4}{\tilde kk}}
            \Bigl(m\HH_2+\mbar \bar \HH_2\Bigr)\Bigr]\ \ .
\label{superpf}
\ee
The bosonic currents and associated fermions are given by
\bbb
       k^1\pm ik^2 &=&
            \psi^{\pm}\, \exp\Bigl[\pm i\sqrt{\txfc{2}{k}}
                    \Bigl(\YY-\sqrt{\txfc{2}{\tilde k}}\,\HH_2\Bigr)\Bigr]
            \nnmb \\[0.2cm]
       \chi^1\pm i\chi^2 &=&
            \sqrt{k}\:\exp\Bigl[\pm i\Bigl(\sqrt{\txfc{2}{k}}\,\YY
            \,+\,\sqrt{\txfc{\tilde k}{k}}\,\HH_2\Bigr)\Bigr]\ ,
\eee
and the supercurrent is written as
\be
       \sqrt{k}\:G \,=\, \psi^{+}e^{-i\sqrt{k/\tilde k}\,\HH_2}
            \,+\,\psi^{-}e^{+i\sqrt{k/\tilde k}\,\HH_2}
            \,+\,\sqrt{2}\,\chi^3\,i\partial\YY\ \ .
\ee
Similarly one may write $X$ and $H_1$ of the $SL(2,\IR)$
supersymmetric WZW model in terms of bosons $\XX$, $\HH_1$
of the analogous superparafermionic construction for $SL(2,\IR)$.
The advantage of the superparafermion description is that
it makes manifest the $\IZ_k$ symmetry of the supersymmetric $SU(2)$
WZW model, which acts only on the boson $\YY$ which
bosonizes the (total) current $J^3_{\sst SU(2)}$
(whereas the bosonic parafermions have a $\IZ_{\tilde k}$ symmetry
acting on $Y$).  Furthermore the superparafermion operators
are by construction primary fields of the superVirasoro algebra.
Thus the twist operators
for the superstring on the orbifold $(\ads_3\times S^3)/\IZ_N\times \NN$
(where $N$ divides $k$) will be written in terms of the
superparafermions ${\widehat{\Psi}}_{jm\mbar}^{\sst SL(2)}$
and ${\widehat{\Psi}}_{jm\mbar}^{\sst SU(2)}$.

\subsubsection*{The \(\left(AdS_{3}\times S^{3}
\right)/\mathbb{Z}_{N}\times T^{4}\) orbifold}
We now consider the superstring on the orbifold \(\left(AdS_{3}\times S^{3}
\right)/\mathbb{Z}_{N}\times T^{4}\,\). The twist vertex operators for
the superstring which have the proper monodromy with respect to the currents
and have a single valued OPE with the worldsheet supercurrent \(G(z)\)
are constructed as follows. The twist operators on
\(SL(2,\mathbb{R})/\mathbb{Z}_{N}\) are given by
the superparafermions
\be
      \Omega^{\sst SL(2)}_{q}\,=\,{\hat\Psi}^{\sst SL(2)}_{j_q,j_q,j_q}\,=\,
	\Psi^{\sst SL(2)}_{j_q,j_q,j_q}\,
	\exp\left[i\sqrt{\txfc{k}{\tilde k}}\;\txfc{q}{N}
	\left(\HH_1+\bar \HH_1\right)\right]\ ;
\ee
here again $j_q=kq/2N$, and $\tilde k=k-Q$.%
\footnote{Note however the shift in notation -- in the supersymmetric
case the order of the cyclic orbifold group $N$ divides the level
$k$ of the {\it total} current, rather than the bosonic
level $\tilde k$.  Hence one must suitably modify the formulae
\pref{slpfdim}, \pref{supfdim} for the parafermion dimensions.}
Similarly the twist operators on \(SU(2) /\IZ_{N}\) are given by
\be
      \Omega^{\sst SU(2)}_{q}\,=\,{\hat\Psi}^{\sst SU(2)}_{j_q,j_q,j_q}\,=\,
	\Psi^{\sst SU(2)}_{j_q,j_q,j_q}\,
	\exp\left[i\sqrt{\txfc{k}{\tilde k}}\;\txfc{q}{N}
	\left(\HH_2+\bar \HH_2\right)\right]\ .
\ee
The dimension of both the superparafermions is
$h(\Omega_q)=q/2N$;
thus the twist operators on the \(\left(AdS_{3}\times S^{3}
\right)/\mathbb{Z}_{N}\times T^{4}\) orbifold are
\be
\Omega^{\rm orb}_{q}\,=\,\Omega^{\sst SL(2)}_{q}\:\Omega^{\sst SU(2)}_{N-q}
	\quad,\qquad q=1,\ldots,N-1\ ;
\label{twistmoduli}
\ee
each of these states has vanishing spacetime energy $\LL_0$
and $S^3$ angular momentum $\TT_0^3$.
The condition for primary fields
to be invariant under the action of the orbifold identification
\be
\exp\left[\,i\left(2\pi /N\right)\left(L_{\sst SL(2)}-L_{\sst
SU(2)}\right)\,\right]
\left|\Phi^{\sst SL(2)}_{jm\bar{m}}\,\Phi^{\sst
SU(2)}_{j'm'\bar{m}'}\right\rangle\,=\,
\left|\Phi^{\sst SL(2)}_{jm\bar{m}}\,\Phi^{\sst
SU(2)}_{j'm'\bar{m}'}\right\rangle
\ee
implies the condition \((m-\bar{m})-(m'-\bar{m}')\in N\mathbb{Z}\,\).
Here the generators of rotation are given by
\be
L_{\sst SL(2)}\,=\,J^{3}_{0}-\bar{J}^{3}_{0}\hspace{2cm}
L_{\sst SU(2)}\,=\,K^{3}_{0}-\bar{K}^{3}_{0}\ .
\ee
Consider the OPE of the twist vertex operators with the spin fields used to
construct the spacetime supercharges
\be
\Omega_{q}^{\sst ORB}(z)\:S_{\alpha}(0)\,\sim\,
z^{\epsilon_{2}/2}\,z^{q/N\left(\epsilon_{1}-\epsilon_{2}\right)/2}\:
:\!\Omega_{q}^{\sst ORB}(z)\:S_{\alpha}(0)\! :\quad;
\ee
this implies \(\epsilon_{1}=\epsilon_{2}\) and thus, from~\pref{brst}
above, \(\epsilon_{3}=1\,\). Thus only 4 `left-moving' supercharges
survive the orbifold projection out of the original 8 on
\(AdS_{3}\times S^{3}\times T^{4}\,\)
(and 4 more from the right-movers).  The associated spin fields can
be indexed by the \(i\partial H_{1}\) and \(i\partial H_{4}\) charges
\be
S_{\epsilon_{1}\epsilon_{4}}\,=\,
e^{\frac{i}{2}\left(\epsilon_{1}(H_{1}+H_{2})+
H_{3}+\epsilon_{4}(H_{4}+H_{5})\right)}\ .
\ee
Note that the twist operators \pref{twistmoduli}
commute with half of the supersymmetry generators,
thus the corresponding states (and their fractional spectral flows)
are BPS.
%
%
\subsubsection*{The supersymmetric (fractional) spectral flow operator
\label{supfracspecflow}}
The integer spectral flow operator introduced in~\cite{M+O} was
extended to the superstring in~\cite{AGS}. Express the total currents
\(J^{3}\) and \(K^{3}\) in terms of bosons \(\XX\) and \(\YY\,\)
\be
J^{3}\,=\,-\sqrt{k/2}\,\partial \XX \hspace{2cm}
K^{3}\,=\,i\sqrt{k/2}\,\partial \YY
\ee
The holomorphic part of the integer spectral flow operator for the
superstring is
\be
t_{w}\,=\,e^{w\,\sqrt{k/2}\,\left(\XX+i\YY\right)}
\label{specflowop}
\ee
Note that \(h(t_{w})=0\) and that \(t_{w}\) is mutually local with
respect to the spacetime supercharges.
Integer spectral flow
$\OO_{\rm\sst BPS}\rightarrow t_w\OO_{\rm\sst BPS}$, $w\in\IZ$,
extends the range of 
BPS operators~\pref{untwistops}
beyond the window $\hf\le j< \frac{k-1}{2}$
of discrete series highest weight representations.
The orbifold again admits fractional spectral flow
under~\pref{specflowop} with $w=p/N$, $p\in\IZ$,
describing oscillating strings of the sort depicted in
figure~1.
The BPS single particle spectrum of the orbifold thus consists of
the surviving untwisted sector
supergravity states~\pref{untwistops}
(\ie\ those with $(m-\mbar)-(m'-\mbar')\in N\IZ$),
together with their
fractional spectral flows by~\pref{specflowop} (with $w=p/N$);
furthermore there are the twist ground states~\pref{twistmoduli}
and their fractional spectral flows.

The continuous representations $\CC_j^\alpha$ of $SL(2,\IR)$
spin $j=\half+is$
appear in the spectrum of the $\ads_3\times S^3\times T^4$ theory
in nonzero spectral flow sectors, where they describe
a continuum of long strings moving in or out from the
boundary of $\ads_3$ with radial momentum $s$.
One might think that the $\IZ_N$ orbifold identification lowers
the threshold of the continuum of long string states
by a factor of order $N$,
since a long string need only wind part of the way around
the $\ads_3$ angular direction.

From the analysis of~\cite{M+O,AGS}, a continuous representation
of spin $j=\hf+is$ can be spectrally flowed $w$ units
and then used to dress a physical vertex operator
$e^{-\varphi}\Phi_{jm}^w\OO_{\rm int}$ (where $\OO_{\rm int}$
is a vertex operator in the remaining $S^3\times T^4$
sigma model), for which the mass shell condition is
\be
     -\frac{j(j-1)}{k} - wm - \frac{k}{4}w^2
	+h_{\rm int}-\half = 0\ ;
\ee
the spacetime energy of this state, as defined on the covering space, is then
\be
     \LL_0=J_0^3=\frac{k}{2}w+m
	=\frac{kw}4+\frac1w\left(\frac{1+4s^2}{4k}
	+h_{\rm int}-\half \right)\ .
\label{thresh}
\ee
Before the orbifold, only $w\in\IZ$ is allowed, and the
threshold of the continuum occurs for $w\!=\!1$,
$s\!=\! 0$, and $h_{\rm int}\!=\!\hf$ (due to chiral GSO),
at $\LL_0=\frac{k^2+1}{4k}$.
The orbifold allows fractional spectral flow
$w\!=\!\frac pN$,
provided that one spectrally flows an amount $-p/N$ in $SU(2)$.
This flows $h_{\rm int}=\hf$ to
$h_{\rm int}=\hf+\frac k4 w^2$ according to the $SU(2)$
version of~\pref{spfl};
the spacetime energy of this state becomes
\be
     \LL_0=\frac{kw}2+\frac1{4kw}\ .
\ee
For $w=1/N$, the energy is indeed reduced by
an amount of order $N$.

\subsection{The $\ads_3/\IZ_N$ Orbifold of the Superstring}

The orbifold of $\ads_3$ alone breaks supersymmetry
in the superstring.  The twist operators are the
superparafermions ${\hat\Psi}^{\sst SL(2)}_{j_q,j_q,j_q}$
of $SL(2,\IR)$.  Their dimension is $kq/2N<1/2$ and thus
the NS sector twist operators describe tachyonic excitations
of the orbifold point; the ground state is unstable.
The twisted sector RR ground state operators take the form
\be
      e^{-\varphi/2}{\hat\Psi}^{\sst SL(2)}_{j_q,j_q,j_q}\,
	\exp\Bigl[\epsilon_1\Bigl(\hf\sqrt{\txfc{2}{k}}\,\XX
	+\txfc{i}{2}\sqrt{\txfc{\tilde k}{k}}\,
		\Bigl(\txfc{k}{\tilde k}-\txfc{2q}{N}\Bigr)\HH_1
	+\txfc{i}{2}\,H_2\Bigr)+\txfc{i}{2}\,H_3
	+\epsilon_4\txfc{i}{2}\bigl(H_4+H_5\bigr)\Bigr)\Bigr]
\ee
where we have suppressed the right-moving free bosons.
These are massless gauge fields coupling to the conserved
RR charge of fractional D-branes (see section~\pref{dbranesec}).
At the orbifold point, the vacuum is unstable, but the
fractional D-branes couple to K-theoretic topological charges;
the D-branes themselves are stable.
As mentioned in the introduction,
if the configuration is to decay to some superposition of
states in the standard $\ads_3\times S^3$ string theory,
the fate of these charges must be understood.


\section{\label{dbranesec}D-branes}

Another class of excitations of the conical defect
are fractional D-branes.  D-branes at an orbifold
singularity $\IR^n/\Gamma$
are classified by the representations
of $\Gamma$, which characterize the orbifold action
on Chan-Paton structure~\cite{douglasmoore,gijo,bcd,ddg}.
Branes in a given irreducible representation are pinned
to the orbifold point; however, assembling irreps into
the regular representation, a moduli space develops
that allows them to be moved off the orbifold point --
in the case of $\IZ_k$, one needs $k$ fractional D-branes to
put one brane on each leaf of the covering space
away from the origin.

In flat spacetime, the tension of a $Dp$-brane
is $\mu_p=(g_s\lstr^{p+1})^{-1}$; however, in the presence
of the nontrivial background \pref{GandB}, the energetics
is modified due to binding energy with the background
fundamental strings and NS fivebranes~\cite{larsmart}.
For example, in the IIB theory, the left-moving
energies of D-branes wrapping the $\T^4$ are
\be
      {\LL}_0 =
            {1\over 4k }\sum_{i=1}^4
	\Bigl({w^{\sst D3}_i\sqrt{v_4}\over r_i}
            +w^i_{\sst D1}\frac{r_i}{\sqrt v_4}\Bigr)^2\ .
\label{Awrapenergy}
\ee
Here
$w_i^{\sst D3}=\frac16\epsilon_{ijkl}w_{\sst D3}^{jkl}$
is the winding charge of D3-branes, and $w^i_{\sst D1}$ is
the wrapping charge of D1-branes; $r_i$ are the radii
of a rectangular $\T^4$, and $v_4=r_1r_2r_3r_4$.
We can then orbifold $\ads_3\times\S^3$ by $\IZ_N$,
and passing to the IIA theory to turn odd branes into
even ones, place fractional branes along the orbifold
singularity, which is extended along one direction
of the $\S^3$; similar D-branes stretched across
a great circle of $S^3$ have been considered
recently in~\cite{malmooseib}.
The size of the $\S^3$ is $\sqrt k$
in string units, so \pref{Awrapenergy} should be
multiplied by $\sqrt k$.  Similarly, in the IIB
theory one has D-branes that lie along the one
dimension of the orbifold singularity in $\S^3$ and
also wrap zero, two, or all four directions
of $\T^4$
\be
      {\LL}_0 =
            {1\over 4\sqrt k }\Bigl(\frac{w_{\sst D1}}{\sqrt{v_4}}
		+w_{\sst D5}\sqrt{v_4}\Bigr)^2+
            {1\over 4\sqrt k }\sum_{i<j}
            \Bigl({^*w}^{\sst D3}_{ij}{\sqrt{v_4}\over r_ir_j}
            +w^{ij}_{\sst D3}\frac{r_ir_j}{\sqrt v_4}\Bigr)^2\ .
\label{Bwrapenergy}
\ee
Assembling fractional branes into regular representations
does not result in a true moduli space for the brane in
$(\ads_3\times S^3)/\IZ_N$; due to the gravitational redshift of
anti-de Sitter space, it costs energy to move an object away from
the origin.  However, the associated energy cost decreases from
string scale for fractional branes to the scale set by the $\ads$
radius of curvature.


\section{\label{modsec}Moduli Space}

In appropriately scaled variables, in the limit $k\to \infty$
the sigma model tends to flat spacetime.
Consequently, the supersymmetric orbifold
of the previous section degenerates to
$\IR^{1,1}\times(\IR^4/\IZ_N)$.
String theory on the orbifold $\IR^4/\IZ_N$ has $4(N-1)$ moduli
corresponding to the resolution of the orbifold singularity
(\cf\ \cite{aspinwall,abfgz}).
The resolution by blowing up inserts $N-1$
two-spheres.  The moduli parametrize the hyperK\"ahler
blowup modes together with the B-flux through the two-spheres;
in particular,
the orbifold point corresponds to a set of collapsed two-spheres
each threaded by a half unit of $B$-flux~\cite{aspinwall}.

   From the construction of the twisted sector spectrum,
we see that the $\ads_3\times\S^3/\IZ_N$ orbifold also has
a set of $4(N-1)$ massless modes corresponding to the operators
\pref{twistmoduli}.
The orbifold effectively acts only on the coset theory
$\frac{SL(2,\IR)}{U(1)}\!\times\! \frac{SU(2)}{U(1)}$,
which is an $\NN=(4,4)$ worldsheet superconformal field theory.
The $\NN=4$ algebra \cite{egutao}
is generated by the stress tensor $T(z)$,
four supercurrents $G^{a\alpha}(z)$, and the $SU(2)$ R-symmetry
current $J^{\alpha\beta}(z)$
(and corresponding antiholomorphic currents);
the supercurrents additionally
transform as a doublet under a global $SU(2)_l$ (the $a$ index).
The $N-1$ moduli multiplets from the twisted sector
transform as $(2,2)$ under this
global $SU(2)_l\times SU(2)_r$ symmetry of the small
$\NN=(4,4)$ superconformal algebra,
and are R-symmetry singlets.
Since the $N-1$ quartets of massless fields all preserve $\NN=4$
worldsheet supersymmetry, all these deformations are
exactly marginal (the trace of the stress tensor lies in
the same supermultiplet as the anomaly in the $SU(2)$ R-symmetry,
which is not renormalized).  Four additional massless multiplets
arise from the untwisted sector and
are universal, being obtained by
worldsheet spectral flow~\cite{schwimmerseiberg}
from the identity sector (see~\cite{nahm} for a recent review).
These are the spin one-half representations of
the $SU(2)$ R-symmetry current algebra, associated in spacetime
to the supergravity multiplet.  However, this last
multiplet is not part of the moduli space of the orbifold --
the expectation value of the zero momentum supergravity
modes are not moduli on a noncompact space.
All told, the $4(N-1)$ massless fields parametrize
the moduli space $O(4,N-1)/O(4)\times O(N-1)$.
It would be interesting to know whether there are
any global identifications (other than the integer periodicity
in the $B$ flux).

For $k\gg N\gg 1$, the orbifold looks locally very much
like $\IR^{1,1}\times (\IR^4/\IZ_N)\times \T^4$,
and an approximate description of the resolved orbifold singularity
as an ALE space is appropriate (up to distances of order $\sqrt{k}\lstr$
from the orbifold point).  The metric on
the ALE space is
\bbb
      ds^2 &=& V^{-1}(\vec x)(d\tau+\vec\omega\cdot d\vec x)^2
	+V(\vec x)d\vec x\cdot d\vec x	\nonumber\\
      V(\vec x) &=& \sum_{i=1}^{N} |\vec x-\vec x_i|^{-1}\ ,
\label{alemetric}
\eee
where $\vec\nabla V=\vec\nabla\times\vec\omega$.
This should accurately describe the vicinity of
the blowup when the deformation is not too far from the orbifold point
in moduli space (which is $\vec x^i=0$ for all $i$).
The $\vec x^i$ parametrize the three metric deformations
of the collapsed spheres,
and can roughly be thought of as the locations of the poles
of the homology two-spheres of the resolved manifold
(the fourth modulus being the $B$-flux through these
homology two-spheres).
These deformations might be thought of
as certain kinds of breathing modes of the object in $\ads_3$.
It is interesting that the object can have a large
number of internal excitations, related to its mass.

\subsubsection*{An aside on related models}

The orbifold action by the maximal discrete
symmetry $\IZ_k$ on the
$\bigl(SL(2)/U(1)\bigr)\times \bigl(SU(2)/U(1)\bigr)$
sigma model was considered in a related context in~\cite{oogurivafa}.
The first factor $\bigl({SL(2)}/{U(1)\bigr)}$ is the Euclidean 2d
black hole or `cigar' sigma model of~\cite{wittenbh,msw},
and also the theory of $SL(2,\IR)$ superparafermions;
the second factor $\bigl({SU(2)}/{U(1)}\bigr)$ is the
the $X^{k}$ Landau-Ginsburg theory~\cite{martlg,vafawarner},
or equivalently the $SU(2)$ superparafermion theory.%
\footnote{The monomial $X^q$ of the Landau-Ginsburg field
is the superparafermion ${\hat\Psi}^{\sst SU(2)}_{j_q,j_q,j_q}$.}
The orbifold
$\bigl[\bigl(SL(2)/U(1)\bigr)\times
\bigl(SU(2)/U(1)\bigr)\bigr]/\IZ_k$
was shown in~\cite{oogurivafa,givkutone,givkuttwo} to give a CFT
with a target space which is asymptotically
$\S^3\times \IR$, corresponding to the throat geometry
of $k$ fivebranes.  It was argued in these works that
the target spacetime is algebraically described by the
deformed $\IZ_k$ singularity
\be\label{Anspace}
      X^k+Y^2+Z^2=\mu\ .
\ee

What is the relation among these various theories?
The operators/states of the $SU(2)$ WZW model can be decomposed into
a parafermion part and a free field part arising from the
bosonization of $J^3$:
\be\label{pfops}
      \Phi^{\sst\rm SU(2)}_{jm\mbar}={\hat\Psi}^{\sst SU(2)}_{jm\mbar}\;
	\exp\Bigl[i\sqrt{\txfc 2k}\Bigl(n\YY+\bar n\bar\YY\Bigr)\Bigr]
\ee
with $m=n$, $\mbar=\bar n$, and $m,\mbar=-j,...,j$.
On the other hand, the tensor product theory
$\left({SU(2)}/{U(1)}\right)_{\rm pf}\times U(1)_{\rm circ}$
consists of the operators on the RHS of \pref{pfops},
with $m,\mbar$ independent of $n,\bar n$,
and $(m-\mbar)\in k\IZ$ (and similarly for $n-\bar n$).
The parafermion operators respect a
$\IZ^{\rm pf}_k\times{\tilde{\IZ}}^{\rm pf}_k$
symmetry, under which the operators carry the
quantum numbers $(l,l')=(m+\mbar,m-\mbar)$ mod $k$~\cite{zamfat}.
Similarly considering a
$\IZ^{\rm circ}_k\times{\tilde{\IZ}}^{\rm circ}_k$
subgroup of the $U(1)\times U(1)$ symmetry of the free boson,
the $SU(2)$ WZW theory is the orbifold of the
parafermion times $U(1)$ theory by the diagonal vectorlike
$\IZ'_k=(\IZ^{\rm pf}_k\times\IZ^{\rm circ}_k)_{\rm diag}$
\be\label{pforb}
      SU(2)=\left[\left(\frac{SU(2)}{U(1)}\right)_{\rm pf}
	\times U(1)_{\rm circ}\right]/\IZ'_k
\ee
under which the states \pref{pfops} carry $\IZ'_k$ charge
$(m+\bar m-n-\bar n)$.
The orbifold by $\IZ'_k$ sets $m=n$, $\mbar=\bar n$;
the twisted sectors relax the condition $m-\mbar\in k\IZ$
to $m-\mbar\in\IZ$.
The relation between the $SL(2,\IR)$ WZW model and
the $\frac{SL(2,\IR)}{U(1)}\times \IR$ cigar plus time
background is not as straightforward; the parafermion states
have spin $j$ taking continuous values $0\le j<\frac{k-1}{2}$.
Nevertheless, the $\IZ_k$ symmetry of~\pref{orbident}
(for $k=N$) acts on the embedded parafermion model
in precisely the same way as in~\cite{oogurivafa,givkutone,givkuttwo}.

There are thus two ways of getting at the orbifold
theory we are discussing in this paper.  Starting with
the target
\be\label{startpt}
      \left({SL(2)}\right)_{\rm WZW}
      \times\left(\frac{SU(2)}{U(1)}\right)_{\rm pf}
	\times U(1)_{\rm circ}\times T^4\quad ,
\ee
one may construct $\ads_3\times S^3$
by orbifolding by $\IZ'_k$
and then performing the orbifold by the $\IZ_k$
of \pref{orbident}, which acts on the embedded
$SU(2)$ and $SL(2)$ parafermions of the respective WZW models.
Alternatively, one can first orbifold by this latter $\IZ_k$,
to make a variant of the CHS model
as in~\cite{oogurivafa,givkutone,givkuttwo},
and then quotient by the first symmetry $\IZ'_k$.

There are correspondingly
a couple of ways to view the deformation along
the moduli space of the orbifold singularity.
One is \pref{alemetric}, which gives an approximate
description of the geometry near the orbifold singularity
for not too large a blowup.
On the other hand, for $k=N\gg 1$,
the description of~\cite{oogurivafa,givkutone,givkuttwo}
in terms of Landau-Ginsburg models is perhaps more appropriate,
and the geometry near the object is stringy.

The division of the target space into superparafermion
and $U(1)$, as in~\pref{startpt}, is the starting point
for a more general orbifold using the
construction of~\cite{AGS}.  There, spacetime CFT vacua
preserving at least eight supersymmetries are built
using $c=15$ worldsheet superconformal field theories
of the form
\be
     \ads_3\times U(1)_k\times \left(\frac{\MM}{U(1)}\right)\ ,
\ee
where $\MM/U(1)$ is $\NN=2$ supersymmetric on the
worldsheet, and the subscript $k$ is the `level' of
the $U(1)$ supercurrent algebra (the radius squared of
the target space circle in units of the self-dual radius).
In this paper we have concentrated on perhaps the simplest example
$\MM/U(1)=SU(2)/U(1)\times T^4$, but clearly one may
generalize.  A $\IZ_N$ orbifold preserving half
the supersymmetry may be constructed along
the lines of sections~\ref{orbsec} and~\ref{suporbsec},
provided $N$ divides $k$.


\section{\label{virsec}The Spacetime CFT}

There is an intimate relation between the asymptotic $\ads$
geometry of the string background and the conformal invariance
of its dual CFT description~\cite{brohen,maldacft,stromads,GKS}. In
the orbifold constuction described above, the quotient
space description of the theory inherits a subalgebra of the
$\NN=4$ superconformal algebra on the covering space which is
itself a twisted $\NN=4$ superconformal algebra.
The central charge of the quotient space superVirasoro algebra
\(c=N\tilde{c}\,\), where \(\tilde{c}\) is the central charge that
appears in the covering space description, is consistent with what is
expected in gravity on asymptotically \(\ads\) spacetimes. A factor
of \(1/N\) appears in \(\tilde{c}\) which is due to the
fact that gravity as described on the covering space has a Planck scale
that is \(N\) times larger than that on the quotient space.
The orbifold geometry can be seen to be the fractional spectral flow
in the spacetime superconformal field theory of a Ramond vacuum state which
carries \(1/N\) of the maximal allowed
\(\TT^{3}_{0}\) charge of the Ramond ground states.
Note that in this section, when we discuss spectral flow,
we are referring to the flow of states in the spacetime
superconformal field theory.

\subsection{The Spacetime Superconformal Algebra}

The spacetime conformal field theory of the onebrane-fivebrane system
is a representation of the $\NN=4$ superVirasoro algebra~\cite{egutao}%
\footnote{At the particular point in the moduli space realized
by the perturbative worldsheet description, the spacetime
CFT is singular; the representation theoretic aspects
of this situation are not well understood.  We thank
D. Kutasov for emphasizing this point.}. This spacetime algebra
on the covering space \(AdS_{3}\times S^{3}\times T^{4}\,\) is as follows
\bbb
     \com{\tilde{\LL}_n}{\tilde{\LL}_m} &=& (n-m)\,\tilde{\LL}_{n+m}+
           \txfc{\tilde{c}}{12}\,n^3\,\delta_{n+m}
	\nnmb\\[.1cm]
     \com{\tilde{\TT}^i_n}{\tilde{\TT}^j_m}&=&
	i\,\epsilon^{ijk}\,\tilde{\TT}^k_{n+m}
	+\txfc{\tilde{c}}{12}\,n\,\delta^{ij}\,\delta_{n+m}
	\nnmb\\[.1cm]
     \com{\tilde{\LL}_n}{\tilde{\TT}^i_m} &=& -\,m\,\tilde{\TT}^i_{n+m}
     \label{superviralg}\\[.1cm]
     \acm{\tilde{\GG}^{*\alpha}_{r}}{\tilde{\GG}^{\beta}_{s}} &=&
	2\delta_{\alpha\beta}\,\tilde{\LL}_{r+s}
	-2(r-s)\sigma^{i}_{\alpha\beta}\,\tilde{\TT}^i_{r+s}
	+\txfc{\tilde{c}}{12}\,4r^2\,\delta_{\alpha\beta}\,\delta_{r+s}
	\nnmb\\[.1cm]
     \com{\tilde{\TT}^i_n}{\tilde{\GG}^{\alpha}_r} &=& \txfc{1}{2}\,
	\sigma^i_{\beta\alpha}\,\tilde{\GG}^{\beta}_{n+r}\
	\nnmb\\[.1cm]
     \com{\tilde{\LL}_n}{\tilde{\GG}^{\alpha}_r} &=&
           -\,(r-\txfc{1}{2}\,n)\,\tilde{\GG}^{\alpha}_{n+r}
     \ \ . \nnmb
\eee
All other (anti-) commutators either vanish or are implied by
the relation \(\,(\tilde{\GG}^{\alpha}_{r})^{\dgr}=
\tilde{\GG}^{*\alpha}_{-r}\,\). The moding of the supercharges
is that of the NS sector associated
with global $\ads_3\times S^3\,$. That is \(r\in \mathbb{Z}+1/2\,\) for
\(\tilde{\GG}^{\alpha}_{r}\,\) and \(n\in \mathbb{Z}\,\) for
$\tilde{\TT}^\pm_n\,$. The central terms appearing here
differ slightly from standard CFT conventions
due to the $\ads_3$ convention that the $SL(2,\IR)$
invariant vacuum state has energy
$\tilde{\LL}_0=-\frac{\tilde{c}}{24}$ rather than zero.
Expressions for the generators of this algebra in terms of
the fields of the sigma model on (Euclidean) $\ads_3\times S^3$
have been given in~\cite{GKS,dort,kutseib}.  The rule of
thumb is that any (anti)holomorphic algebra on the
worldsheet is related to a corresponding algebra in spacetime;
roughly, long strings near the boundary of $\ads_3$
have the spacetime algebra pulled back onto the worldsheet.

The $\IZ_N$ orbifold of spacetime projects
the generators~\pref{superviralg} onto the subalgebra
which commute with the $\IZ_N$ action generated by
\be
      \exp[\coeff{2\pi i}{N}(\tilde{\LL}_0-\tilde{\bar\LL}_0
      -\tilde{\TT}^3_0+\tilde{\bar\TT}^3_0)]\ .
\label{ZNact}
\ee
For example, of the Virasoro and R-current generators $\tilde{\LL}_n$
and $\tilde{\TT}^3_n$, one keeps only those with $n\in N\IZ$;
similarly $\tilde{\GG}^{\pm}_r$ survive for $r\pm\hf\in N\IZ$
(so that $\tilde{\GG}^{\pm}_{\mp 1/2}$ are the surviving supersymmetries),
and we also keep $\tilde{\TT}^\pm_n$ for $n\pm 1\in N\IZ$.
Here we have defined $\tilde{\GG}^{+}_r\equiv\tilde{\GG}^{1}_r$ and
$\tilde{\GG}^{-}_r\equiv\tilde{\GG}^{2}_r$ to reflect the
$\tilde{\TT}^3_0$ eigenvalue. We similarly
define $\tilde{\GG}^{*+}_r\equiv\tilde{\GG}^{*2}_r$ and
$\tilde{\GG}^{*-}_r\equiv\tilde{\GG}^{*1}_r\,$.
Note that the effect of the orbifold on the spacetime
CFT is {\it not} simply to project onto the states
that are invariant under~\pref{ZNact}; this would allow
multiparticle states invariant under~\pref{ZNact} built out
of particles that were not individually invariant under~\pref{ZNact}
(for instance, operators in the Virasoro enveloping algebra
built out of products of Virasoro raising operators whose
total level is a multiple of $N$ but whose component raising
operators have levels that are not multiples of $N$).
Such states are not present in the orbifold.

In the twisted sector of the orbifold, no new (anti)holomorphic
currents appear on the worldsheet, and therefore one
expects that no new symmetries of spacetime will appear.
The $\IZ_N$ projected superVirasoro algebra
is then the full set of symmetries of the spacetime. We may define the
following spacetime generators associated with a twisted superconformal
symmetry on the quotient space
\bbb
     &&\LL_n = \txfc{1}{N}\,\tilde{\LL}_{nN} \nnmb\\[0.1cm]
     &&\TT^3_n = \tilde{\TT}^3_{nN}
\label{newgens}\\[0.1cm]
     &&\TT^{\pm}_{n\mp\frac{1}{N}} = \tilde{\TT}^\pm_{nN\mp 1} \nnmb \\[0.1cm]
     &&\GG^{\pm}_{n\mp\frac{1}{2N}} = \txfc{1}{\sqrt{N}}\,
	\tilde{\GG}^{\pm}_{nN\mp\frac{1}{2}} \quad . \nnmb
\eee

\noindent Taking $\,c=N\tilde{c}\,$, these generators
satisfy the following algebra

\bbb
     \com{\LL_n}{\LL_m} &=& (n-m)\LL_{n+m}\,+\,\txfc{c}{12}\,n^3\,\delta_{n+m}
	\nnmb\\[.1cm]
     \com{\TT^3_{n}}{\TT^3_{m}} &=& \txfc{c}{12}\,n\,\delta_{n+m}
	\nnmb\\[.1cm]
     \com{\TT^3_{n}}{\TT^{\pm}_{m\mp\frac{1}{N}}} &=&
           \pm\,\TT^{\pm}_{n+m\mp\frac{1}{N}}
	\nnmb\\[.1cm]
     \com{\TT^{+}_{n-\frac{1}{N}}}{\TT^{-}_{m+\frac{1}{N}}} &=&
           2\,\TT^{3}_{m+n}\,+\,2\,\txfc{c}{12}
	\left(n-\txfc{1}{N}\right)\delta_{n+m}
	\nnmb\\[.1cm]
     \com{\LL_{n}}{\TT^3_{m}} &=& -\,m\,\TT^3_{n+m}
	\nnmb\\[.1cm]
     \com{\LL_{n}}{\TT^{\pm}_{m\mp\frac{1}{N}}} &=&
           -\,(m\mp\txfc{1}{N})\,\TT^{\pm}_{n+m\mp\frac{1}{N}}
	\label{orbalg}\\[.1cm]
     \acm{\GG^{*\pm}_{n\mp\frac{1}{2N}}}
         {\GG^{\pm}_{m\mp\frac{1}{2N}}} &=&
         -2\,(n-m)\,\TT^{\pm}_{n+m\mp\frac{1}{N}}
         \nnmb\\[.1cm]
     \acm{\GG^{*\pm}_{n\mp\frac{1}{2N}}}
         {\GG^{\mp}_{m\pm\frac{1}{2N}}} &=&
         2\,\LL_{n+m}\,\pm 2(n-m\mp\txfc{1}{N})\,\TT^3_{n+m}
         +\txfc{c}{12}\,4(n\mp\txfc{1}{2N})^2\,\delta_{n+m}
         \nnmb\\[.1cm]
     \com{\TT^3_{n}}{\GG^{\pm}_{m\mp\frac{1}{2N}}} &=&
          \pm\,\GG^{\pm}_{n+m\mp\frac{1}{2N}}
          \nnmb\\[.1cm]
     \com{\TT^{\pm}_{n\mp\frac{1}{N}}}{\GG^{\mp}_{m\pm\frac{1}{2N}}} &=&
          \GG^{\pm}_{n+m\mp\frac{1}{2N}}
          \nnmb\\[.1cm]
     \com{\LL_n}{\GG^{\pm}_{m\mp\frac{1}{2N}}} &=&
           -\,(m\mp\txfc{1}{2N}-\txfc{1}{2}\,n)\,
	\GG^{\pm}_{n+m\mp\frac{1}{2N}}\quad .\nnmb
\eee

Thus we can define generators
which satisfy a form of the $\NN=4$
superVirasoro algebra for $ c=N\tilde{c}\,$,
spectrally flowed~\cite{schwimmerseiberg}
by $-1/N$ units from Ramond boundary conditions.

A number of issues arise from the form of the above
quotient space superconformal algebra. One is the question of
the spacetime \(\LL_0\) and \(\TT^3_0\) charges of the orbifold
geometries themselves, in a suitably chosen asymptotic frame, in the
\(SU(1,1|2)\times SU(1,1|2)\) supergravity that (naively)
arises from a Kaluza-Klein reduction
on \(S^{3}\,\). Related to this is the issue of how the above moding
of the supercharges is related to the \(SU(2)\) gauge field of the
extended supergravity and how spectral flow enters the picture.
Another question is the appearance of the factor \(c=N\tilde{c}\) in
the superalgebra and its interpretation in the boundary theories on
the covering and quotient spacetimes. We will answer these questions
in the next two sections but will provide here a simplified
description of the relationship between the covering and quotient
space algebras in the case of an bosonic orbifold of \(\ads_{3}\)
without a corresponding twisting of the \(S^{3}\,\).
That is we consider a metric on the covering space of the form
\be
d\tilde{s}^{2}/k \,=\, -(1+\tilde{r}^{2})\,d\tilde{t}^{2}+
(1+\tilde{r}^{2})^{-1}\,d\tilde{r}^{2}+
\tilde{r}^{2}d\tilde{\phi}^{2} \quad .
\ee
Under the orbifold identification
\(\tilde{\phi}\sim \tilde{\phi}+\txfc{2\pi}{N}\,\) with the following
coordinate transformation
\be
\label{orbsl2crds}
     t=N\tilde{t}\ ,\quad r=\tilde{r}/N\ , \quad  \phi=N\tilde{\phi}
     \quad ,
\ee
we find the following metric on the quotient spacetime
\be
ds^{2}/k \,=\, -(N^{-2}+r^{2})\,dt^{2}+(N^{-2}+r^{2})^{-1}\,dr^{2}+
r^{2}d\phi^{2} \label{orbsl2met} \quad .
\ee
More generally this coordinate transformation defines a map between metrics
which have angular periodicity \(2\pi/N\) and are asymptotically
\(\ads\) on the covering space and metrics
which have angular periodicity \(2\pi\) and are asymptotically
\(\ads\) on the quotient space.
To see that the above form of the relation between the spacetime
generators as defined on the covering space and those defined on
the orbifold is the correct one, we examine the transformation
properties of the worldsheet form of the superconformal generators.
The generators of the asymptotic symmetries of asymptotically
\(\ads_{3}\) spacetimes can be derived from the sigma model
action via a Noether procedure. The generators
of the bosonic Euclidean \(SL(2,\mathbb{R})\times SU(2)\) WZW model will be
discussed briefly here. The Euclidean version of the
\(SL(2,\mathbb{R})\) theory is the
\({\rm H_{3}^{+}}=SL(2,\mathbb{C})/SU(2)\)
coset model.
Diffeomorphisms will, when acting on physical vertex operators, lead
to states that are the same element in the worldsheet BRST
cohomology.
This is not the case for the asymptotic symmetries
of \(\ads_{3}\,\) since these are not globally well defined diffeomorphisms.
In particular, the set of geometries given by the
action of the symmetries on global \(\ads_{3}\) are physically
distinguishable and span the space of all asymptotically
\(\ads_{3}\) spacetimes. Holomorphic contour integral representations
of the operators may be derived for \(\LL_{n}\) and \(\TT^{a}_{n}\). We
express these generators in the \((\tau=it,r,\phi)\) coordinates
in the limit \(r\rightarrow\infty\,\). Taking \(w=\tau+i\phi\) we find
\bbb
\LL_{n} & = & \frac{k}{2\pi i} \oint dz\, e^{-n\,\bar{w}}
\left(r^{2}\,\partial w\,+\,n\,\frac{\partial r}{r}\right) \nnmb \\[0.3cm]
\TT^{a}_{n} & = & \frac{1}{2\pi i} \oint dz\, e^{-n\,\bar{w}}\,K^{a} \quad .
\eee
Where \(K^{a}\) is the holomorphic \(SU(2)\) current.
Using the transformation between covering space and quotient space
variables given above (\(\,r=\tilde{r}/N\,\) and
\(\,w=N\tilde{w}\,\)) it may be seen that
\(\,\tilde{\LL}_{Nn}/N=\LL_{n}\,\) and
\(\,\tilde{\TT}^{a}_{Nn}=\TT^{a}_{n}\,\), in agreement with the expressions
given above for the superconformal generators. For instance, for the
\(SL(2,\mathbb{R})\) invariant NS ground state,
\(\tilde{\LL}_0=-\tilde{c}/24\,\) and
\(\,\LL_0=-c/(24N^{2})=\tilde{\LL}_0/N\,\). Here we have determined
\(\tilde{\LL}_0\) and \(\LL_0\) from the form of the metric. In general
the following metric, in quotient space variables,
describes the conical spacetimes as well as the
BTZ black holes
\bbb
     ds^{2} &=& - N_t^2\,dt^2+N_t^{-2}\,dr^2
	+r^2(d\phi-N_\phi\,dt)^2 \nnmb\\[.1cm]
     N_t^2 &=& \frac{r^2}{\rads^2}-8\lpl M
	+\frac{16\lpl^{2}J^2}{r^2}
\label{btzmet} \\[.1cm]
     N_\phi &=& \frac{4\lpl J}{r^2}\ . \nnmb
\eee
Geometries with $M>0$ are black holes, while those with $M<0$
are conical defects. Using \(c=3\rads/2\lpl\)
we have \(\rads M=\LL_{0}+\bar{\LL}_{0}\) and
\(J=\LL_{0}-\bar{\LL}_{0}\,\).
As will be seen below, this relationship will be modified when we
consider gauge fields arising from the \(S^{3}\) fibration.

\subsection{The Spacetime Charges and Spectral Flow}

In the case of the supersymmetric orbifold considered above, the
charges measured by an asymptotic observer are modified as described
in \cite{Henneaux:1999ib,malmao,bdkr}.
It is first necessary to specify the
relationship between the covering and quotient space coordinates. We
begin with the metric on \(\ads_{3}\times S^{3}\) expressed in covering
space coordinates. Taking the length scales to be identical
\bbb
\label{prodmet}
d\tilde{s}^{2}/k & = & -(1+\tilde{r}^{2})\,d\tilde{t}^{2}+
(1+\tilde{r}^{2})^{-1}\,d\tilde{r}^{2}+
\tilde{r}^{2}d\tilde{\phi}^{2} \nnmb \\[0.2cm]
& & +\cos^{2}\tilde{\chi}\,d\tilde{\theta^{2}}+d\tilde{\chi}^{2}+
\sin^{2}\tilde{\chi}\,d\tilde{\psi}^{2} \quad .
\eee
The \(S^{3}\) coordinate \(\tilde{\psi}\) was given by
\(\phi_{\sst\rm SU(2)}\)
above (see equation \pref{phidef})
and the
\(\ads\) metric takes the form~\pref{GandB} when
\(\tilde{r}=\sinh\tilde{\rho}\,\).
The orbifold is given by the identification
\((\tilde{\phi},\tilde{\psi})\sim(\tilde{\phi}+
\txfc{2\pi}{N},\tilde{\psi}-\txfc{2\pi}{N})\,\). We may define new
coordinates on the quotient spacetime
\bbb
     && t=N\tilde{t}\ ,\quad r=\tilde{r}/N\ ,
     \quad  \phi=N\tilde{\phi}\ \ , \nnmb\\[0.2cm]
     && \theta=\tilde{\theta}\ ,\quad \chi=\tilde{\chi}\ ,
     \quad  \psi=\tilde{\psi}+\tilde{\phi}\ \ ,
\label{orbcoords}
\eee
which may be seen to satisfy \((\phi,\psi)\sim(\phi+2\pi,\psi)
\sim(\phi,\psi+2\pi)\,\). This leads to the quotient space metric
\bbb
\label{orbmet}
ds^{2}/k & = & -(N^{-2}+r^{2})\,dt^{2}+(N^{-2}+r^{2})^{-1}\,dr^{2}+
r^{2}d\phi^{2} \nnmb \\[0.2cm]
& & +\cos^{2}\chi\,d\theta^{2}+d\chi^{2}+
\sin^{2}\chi\,(d\psi-N^{-1}d\phi)^{2} \quad .
\eee
The \(B\) field is of course similarly modified. Note that this choice
of coordinates establishes the charges of the orbifold spacetimes with
respect to the associated asymptotic frame. It will be shown that this
particular choice places the orbifold in the Ramond sector of boundary
CFT. Any other choice that is related by a globally well-defined
coordinate transformation is equivalent but will lead to a different
asymptotic frame with different global charges. The Kaluza-Klein
reduction of the metric and \(B\) fields give rise to a pair of gauge
fields in \(SU(1,1|2)\times SU(1,1|2)\) supergravity.
Following \cite{bdkr}, these are given by
\be
A\,=\,\bar{A}\,=\,-N^{-1}\,\sigma^{3}\,d\phi
\ee
Furthermore the global charges are modified by the
presence of the gauge fields. Let us define the
global charges \(Q^{a}\)
\be
Q^{a}\,=\,\frac{1}{2\pi}\oint\,\txfc{1}{2}\,{\rm tr}[\,\sigma^{a}A\,]
\ee
And similarly for \(\bar{Q}^{a}\,\). In terms of these charge and the
charges \(M\) and \(J\,\), which may be derived similarly by expressing
the Einstein-Hilbert action in terms of gauge fields for a
\(SL(2,\mathbb{R})\times SL(2,\mathbb{R})\) Chern-Simons theory, the
global charges are given by
\bbb
\TT^3_0 & = & \frac{c}{12}\,Q^{3} \nnmb \\[0.2cm]
\bar{\TT}^3_0 & = & \frac{c}{12}\,\bar{Q}^{3} \\[0.2cm]
\LL_0 & = & \txfc{1}{2}(\rads M+J)\,+\,
\frac{c}{24}\,Q^{a}Q^{a} \nnmb \\[0.2cm]
\bar{\LL}_0 & = & \txfc{1}{2}(\rads M-J)\,+\,
\frac{c}{24}\,\bar{Q}^{a}\bar{Q}^{a} \nnmb \ \ \ .
\eee
These relations lead to the charges
\(\,\TT^3_0=\bar{\TT}^3_0=-\frac{c}{12}N^{-1}\,\) and
\(\,\LL_0=\bar{\LL}_0=0\,\).
The orbifold thus lies in the Ramond sector of the spacetime CFT when
charges are defined with respect to the choice of quotient space frame
given by~\pref{orbcoords}. Note that had we chosen the asymptotic
frame \(\,\psi=\tilde{\psi}+(N-1)\tilde{\phi}\,\) we would have found
\(\,A=\bar{A}=(1-N^{-1})\,\sigma^{3}d\phi\,\). This choice identifies the
orbifold spacetimes as being chiral primary states in the
NS sector of the spacetime CFT, with charges
\(\,\TT^3_0=\LL_0+\frac{c}{24}=\frac{c}{12}(1-N^{-1})\,\).%
\footnote{The identification of these coordinate choices with the
R and NS sectors may be seen by expressing a \(2\pi\)
rotation in \(\psi\) in the respective quotient space frames in terms of
group actions on the covering space. The action of the respective
group elements on the supercharges \pref{superqs} will
produce \(-1\) for the NS sector (as it does for global \(\ads_{3}\))
and \(+1\) for the R sector.
}
How does this analysis comport with the conclusion,
based on the relationship between the 
superconformal generators~\pref{newgens}, that
\(\,\LL_0=-c/(24N^{2})=\tilde{\LL}_0/N\,\) and
\(\,\TT^3_0=\tilde{\TT}^3_0=0\,\)? 
These latter values for the charges follow from the
fact that the covering space -- global \(\ads_{3}\) -- has charges
\(\tilde{\LL}_0=-\tilde{c}/24\) and \(\tilde{\TT}^3_0=0\,\). 
The answer is that the superalgebra~\pref{orbalg} corresponds to gauging
away the \(A\) and \(\bar{A}\) fields for each choice of \(N\,\).
This is just the operation of spectral flow in the boundary CFT which
produces the moding of the supercharges seen in~\pref{orbalg}.
Spectral flow induces the shifts
\bbb
\LL_0(\eta) & = & \LL_0\,-\,
\TT^3_0\,\eta\,+\,\frac{c}{24}\,\eta^2 \nnmb \\[0.2cm]
\TT^{3}_{0}(\eta)& = & \TT^{3}_{0}\,-\,\frac{c}{12}\,\eta \ \ .
\eee
The Ramond vacuum state which is the \(\eta=1\)
spectral flow of the NS groundstate thus has
\(\LL_0=0\) and \(\TT^3_0=-c/12\,\).
Starting with a state with charges \(\LL_0=0\) and
\(\TT^3_0=-c/(12N)\,\), which are the charges with respect to the
gauge choice corresponding to~\pref{orbcoords}, a fractional
spectral flow by \(-1/N\) leads to
\be
\LL_0\,=\,-\frac{c}{24N^{2}}\spwd{1.2cm}{and}
\TT^3_0\,=\,0  \ \ \label{orbcharges},
\ee
which are the charges with respect to the superalgbra~\pref{orbalg}. 
Note that it is more consistent to express the charges of the orbifolds
with respect to a particular asymptotic frame which places them in the
same spectral flow sector. The charges~\pref{orbcharges} correspond
to a different gauge choice for each orbifold and thus to a different
choice of (in general only locally asymptotically \(\ads_{3}\))
quotient space coordinate system for each. However, keeping this in
mind, we may choose to express the charges of the orbifolds in terms
of~\pref{orbalg}. Figure~2 shows a plot of the
\(\LL_0\) and \(\TT^3_0\) charges of the
superalgebra~\pref{orbalg} for the \(N=2\) orbifold. The other orbifolds
may be seen to lie along the line given by the
spectral flow of the (\(\LL_0=\TT^3_0-\frac{c}{24}\)) chiral primary states.
Also shown in the plot is the unitarity bound given by
\(\LL_0=(6/c)(\TT^3_0)^{2}-\frac{c}{24}\) that applies to all states
in all of the spectral flow sectors of the superconformal algebra.

\begin{figure}[h]
\label{figtwo}
\begin{center}
\[
\mbox{\begin{picture}(240,180)(0,0)
\put(128,182){\(\LL_0\)}
\put(248,60){\(\TT^3_0\)}
\includegraphics{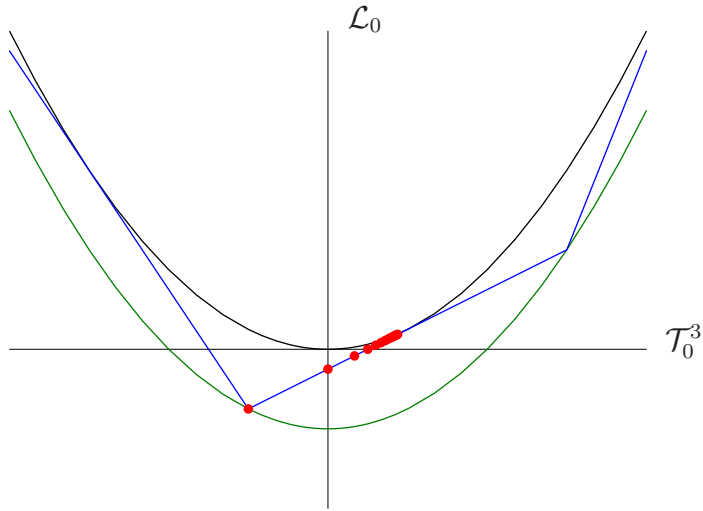}
\end{picture}}
\]
\caption{Charges of the spectral flow of
the chiral primary states (the \textcolor{red}{red} dots)
associated with the orbifold spacetimes
for the sector corresponding to the \(N=2\) orbifold.
The plot shows the unitarity bound
(the circumscribed \textcolor{wtmgreen}{green} parabola)
on all of the states of the
\(\mc{N}=4\) superconformal algebra as well as the more
restrictive bound (the \textcolor{blue}{blue} polygon) that
applies to the sector of the \(N=2\) orbifold. The plot also shows
the lower bound (the inscribed \textcolor{black}{black} parabola)
for the emergence of BTZ black holes.}
\end{center}
\end{figure}

\subsection{The Spacetime Central Charge}

The possibilty for confusion exists over the factor of \(N\) in the
relationship (\(c=N\tilde{c}\)) between the central charge
in the covering space and that in the quotient
spacetime.%
\footnote{In particular, it is possible to draw erroneous conclusions
about the aymptotic growth of the density of states in the spacetime
CFT and consequences for black hole entropy. Evidence for the existence
of such a pitfall is given by statements in the original version of
this paper.}
The issue is seen to appear in the description of low energy gravity on
the respective spacetimes. Ignoring surface terms, consider the
Einstein-Hilbert action on the quotient spacetime
\be
S_{\rm EH}\,=\,\frac{1}{16\pi\lpl}
\int dt\,dr\int_{0}^{2 \pi}d\phi\,\sqrt{-g}\;R[g]
\ee
We are considering metrics which are asymptotically \(\ads_{3}\) on
the quotient space with periodicity \(2\pi\) in the angular
coordinate \(\phi\,\).
We now write this action in the covering space coordinates given by the
transformation~\pref{orbsl2crds}.
Since the metric has periodicity \(2\pi/N\) in the angular
coordinate \(\tilde{\phi}\,\), we may integrate over the entire
covering space while dividing by \(N\) and replacing the quotient
space \(\lpl\) by the corresponding scale on the covering space
\(\tilde{\lpl}=N\lpl\)
\be
S_{\rm EH}\,=\,\frac{1}{16\pi\tilde{\lpl}}
\int d\tilde{t}\,d\tilde{r}\int_{0}^{2 \pi}d\tilde{\phi}\,\sqrt{-\tilde{g}}\;
R[\tilde{g}]
\ee
The asymptotic Virasoro algebra on \(\ads_{3}\) may be derived from
either of these forms of the action with the understanding that there
is a restriction on the space of metrics defined on the covering
space to those that have \(2\pi/N\) periodicity in the angular
coordinate \(\tilde{\phi}\,\). As described above, this will project
out those generators of the algebra which do not respect this
periodicity. The central charge of the Virasoro algebra on the
covering space will be \(\tilde{c}=3\rads/2\tilde{\lpl}=c/N\,\).
Thus the introduction of the factor of \(N\) into the
central charge of the algebra~\pref{orbalg}
corresponds to a return to the central charge
of the quotient spacetime. Another way to see the emergence of the
factor of \(N\) in the central charge on the covering space is to note
that in the formula \(c=6kp\) the integer \(p\) represents the number
of fundamental strings which produce the background.
If there are \(\tilde{p}\) of these on the covering space
then there will be \(p=N\tilde{p}\) on the orbifold since
each of the \(\tilde{p}\) strings will wind \(N\) times around
the quotient spacetime. The same factor of \(N\) may be derived
using the transformation~\pref{orbsl2crds}
from the worldsheet vertex operators associated with the central
charge of the Virasoro algebra derived in~\cite{GKS,dort,kutseib}.

BTZ black holes may be constructed by dumping energy into
the system.  These have an entropy
\be\label{btzent}
      S = \frac{A}{4\lpl}
	= 2\pi\Bigl(\frac{c}6\,\LL_0-(\TT_0^3)^2\Bigr)^{1/2}
	+2\pi\Bigl(\frac{c}6\,\bar\LL_0-(\bar\TT_0^3)^2\Bigr)^{1/2}
\ee
which is invariant under spectral flow.
The BTZ black hole threshold is indicated in Figure~2;
it touches the line of BPS states midway between its
extremes.  The orbifold geometries thus approach the black
hole threshold from below at large $N$.

For the non-supersymmetric orbifold, we may
approach the black hole threshold arbitrarily closely,
since there is no bound on $N$.
For the supersymmetric orbifold we get closest to the
BTZ threshold by taking $N=k$.
The energy below the threshold at this point is of order
$pk/N^2=p/k=g_6^{-2}$ in units of the AdS radius.
One is still a macroscopic distance in configuration
space from the BTZ black hole threshold.

Even though we cannot approach the density of states of
the BTZ black hole in a regime where low-energy supergravity
is reliable, the $N=k$ orbifold does have a large
number of degrees of freedom concentrated in the vicinity
of the orbifold singularity, and it will be interesting
to see to what extent the properties of the orbifold
parallel those of the black hole that arises at higher mass.


\section{\label{dissec}Summary}

We have shown that the $SL(2)$ and $SU(2)$
(super)parafermion theories~\cite{DPL,zamfat,gepqiu}
provide a natural formulation of the $\IZ_N$ orbifold
of string theory on $\ads_3\times S^3$ (times $T^4$ or $K3$).
The spacetime so constructed is a kind of embedding of an
ALE singularity into $\ads_3\times S^3$.
Twisted sector vertex operators are built out of the parafermion highest
weight states, and the spectral flow operation used~\cite{M+O,AGS}
to fill out the single-string spectrum extends to a
fractional spectral flow generating radially oscillating
strings that close only up to the orbifold identification.
The $N-1$ twisted sectors provide $4(N-1)$
new moduli of the worldsheet CFT, analogous to the
hyperK\"ahler blowup modes of an ALE singularity.
These may have an interpretation as describing a
$4(N-1)$ parameter family of states in the spacetime CFT.

We also explored the symmetry algebra of the spacetime CFT
and the associated black hole entropy formula.
The orbifold spacetime is a representation
of a fractionally moded $\NN=(4,4)$ superconformal algebra;
the fractional moding can be obtained by $1/N$ spectral
flow from Ramond boundary conditions.
Flowing back to Ramond boundary conditions, the collection
of orbifolds of various $N$ describe a discrete set of
BPS states in the spacetime CFT, corresponding to
supersymmetric conical defects which approach the
extremal BTZ black hole for $N$ large.

The twisted sector states provide a set of excitations
confined to the orbifold singularity.  There is a macroscopic
number of such excitations when $N$ is large, and
it will be interesting to see to what extent these
excitations mimic the behavior of the black hole states
that appear at higher energy.  Work in this direction
is in progress.

\vskip 2cm
\noindent
{{{\bf Acknowledgments}}}:
We thank
V.~Balasubramanian,
B.~da Cunha,
J.~de Boer,
J.~Maldacena,
A.~Giveon,
E.~Keski-Vakkuri,
D.~Kutasov
and
S.~Ross
for discussions and correspondence.
This work was supported by DOE grant DE-FG02-90ER-40560.

\newpage


\providecommand{\href}[2]{#2}\begingroup\raggedright\endgroup

\end{document}